\documentclass[aps,prd,twocolumn,aps,floatfix,showpacs,showkeys,superscriptaddress,nofootinbib]{revtex4-1}

\usepackage{amssymb}
\usepackage{amsbsy}
\usepackage{amsthm}
\usepackage{amsmath}
\usepackage{breqn}
\usepackage{bm}
\usepackage{braket}
\usepackage{color}
\usepackage{dcolumn}
\usepackage{epsfig}
\usepackage[T1]{fontenc}
\usepackage{graphicx}
\usepackage[colorlinks]{hyperref}
\usepackage{mathtools}
\usepackage{placeins}
\usepackage{xcolor}
\usepackage{cleveref}
\usepackage{subfigure}

\begin{document}
	

\title{Quantum Liang Information Flow as Causation Quantifier}

	\newcommand{\affone}{Department of Physics and Astronomy, University College London, Gower Street, WC1E 6BT London, United Kingdom.}

\author{Bin Yi}
\affiliation{\affone}

\author{Sougato Bose}
\affiliation{\affone}



	\date{\today}
	
	\begin{abstract}
		Liang information flow is a quantity widely used in classical network theory to quantify causation, and has been applied widely, for example, to finance and climate. The most striking aspect here is to freeze/subtract a certain node of the network to ascertain its causal influence to other nodes of the network. Such an approach is yet to be applied to quantum network dynamics. Here we generalize Liang information flow to the quantum domain using the von-Neumann entropy. Using that we propose to assess the relative importance of various nodes of a network to causally influence a target node. We exemplify the application by using small quantum networks.
	\end{abstract}
	
	\maketitle
\paragraph*{Introduction}
The significance of information flow lies not only in communication, but also in its logical implication of causation\cite{hlavavckova2007causality,pearl2000models,spirtes2000causation,bezruchko2010extracting,schreiber2000measuring}. Established in the context of classical physics, the mathematical theory of causality has been widely applied to a variety of disciplines, e.g., climate science\cite{wang2004relation,runge2012quantifying}, network dynamics\cite{sun1994neural,ay2008information,sommerlade2009estimating,timme2014revealing}, neuroscience\cite{pereda2005nonlinear,friston2003dynamic,schelter2006testing,staniek2008symbolic,andrzejak2011characterizing,wu2008detecting}, fiance\cite{marschinski2002analysing,lee2012jumps}, turbulence\cite{tissot2014granger,materassi2014information}. Historically, various measures of classical information flow were proposed\cite{runge2012quantifying,vastano1988information,sun2014causation,schreiber2000measuring,marschinski2002analysing,staniek2008symbolic,duan2013direct}, nonetheless, limitations were pointed out\cite{hahs2011distinguishing,smirnov2013spurious,sun2014causation}. In light of the limitations, Liang and Kleeman established a universally applicable formalism within the framework of classical dynamical systems\cite{san2005information,san2008information,san2014unraveling,san2016information,liang2021measuring,liang2021normalized}. The series of work puts the notion of information flow and causation on a rigorous footing, as Liang(2016)\cite{san2016information} argued: "Information flow and causality can be derived \textit{ab initio}" The formalism has been validated with various benchmark cases\cite{san2016information}, and successfully applied to many realistic problems: glaciology\cite{vannitsem2019testing}, neuro-science\cite{hristopulos2019disrupted}, El Ni\~{n}o-Indian Ocean Dipole relation\cite{san2014unraveling}, precipitation-soil moisture interaction\cite{hagan2019time}, global climate change\cite{stips2016causal}, etc.

The discussion of causality in quantum physics goes back to the paradigmatic Bell experiment\cite{bell1964einstein}. Causal structure places constraints on the correlations that can be generated in any classical hidden variable theories, which quantum physics violates\cite{freedman1972experimental,aspect1982experimental,christensen2013detection,rowe2001experimental,giustina2013bell}. Motivated by the connection between causality and correlations, various attempts have been made to estimate causal influences in certain quantum environments\cite{gachechiladze2020quantifying,henson2014theory,chaves2015information,costa2016quantum,fritz2016beyond,barrett2019quantum,wolfe2021quantum,aaberg2020semidefinite}. The quantification of causal effects in quantum regime shed new light on quantum communication\cite{fitzsimons2015quantum,pisarczyk2019causal}. Furthermore, a information-flux approach was introduced for many-body systems to quantify the influence from a specific element to another, aiming to facilitate the design of quantum processors equipped with large registers\cite{di2007information,di2008deeper}.  In quantum mechanics, correlation functions of Heisenberg picture evolving operators are often used to ascertain casual influences, but one has to be careful that correlations do not imply causation.  

Somewhat counterintuitively, the most straightforward approach to ascertaining causality, for example, one which an experimentalist will naturally employ, namely, to subtract a given component from a network to quantify its influence on other subsystems, remain unexplored. Motivated by that, in this work, we adopt Liang's methodology to establish a formalism of quantum information flow. As opposed to all the approaches mentioned above in the quantum context, here one detaches or freezes a certain subsystem of a network (sender) in order to ascertain its causal influence on other subsystems (target). The change of a target element's von-Neumann entropy, which possess various interpretations\cite{nielsen2002quantum}, then defines the information flow from the sender. When the sending and receiving elements evolve independently, then the information flow measure vanishes.

\paragraph*{Definition:} Consider arbitrary multi-partite system with density state $\rho$, evolving under unitary operator $U(t)=e^{-iHt/\hbar}$, with Hamiltonian $H$. Following Liang's methodology (briefly reviewed in the Supplementary Material[SM]), we decompose the time rate change of von-Neumann entropy of subsystem A, $dS_A/dt$, into two parts: $T_{B\rightarrow A}$, the rate of information flow from subsystem B to A, and $\frac{dS_{A\not{B}}}{dt}$, the entropic evolution rate of subsystem A with influence from B excluded: 
\begin{equation}\label{flow}
T_{B\rightarrow A}=\frac{dS_A}{dt}-\frac{dS_{A\not{B}}}{dt}
\end{equation}
$S$ is von-Neumann entropy given by $S=-\mathrm{Tr}(\sigma \mathrm{log}\sigma)$ for arbitrary density state $\sigma$. $S_{A\not{B}}=S(\rho_{A\not{B}})=S[\varepsilon(t)_{A\not{B}} \rho_A(0)]$, where $\varepsilon(t)_{A\not{B}}$ is a map denoting the evolution of A with B frozen. We will discuss the definition and properties of $\varepsilon(t)_{A\not{B}}$ in the following section. If we consider time evolution as a discrete mapping during interval $\Delta t$, the cumulative information flow is then:
\begin{equation}
\mathbb{T}_{B\rightarrow A}=\int T_{B\rightarrow A} dt=
\Delta (S_A-S_{A\not{B}}) \label{discreteT}
\end{equation}
Note that von-Neumann entropy, therefore the information flow formalism, possess various interpretations\cite{nielsen2002quantum}. Particularly distinct from classical Shannon entropy, von-Neumann entropy quantifies the entanglement within a pure bipartite quantum system. $S_{A\not{B}}$ (or $S_A$) can then be interpreted as the entanglement between A and the rest of the universe with (or without) B frozen. The term $(S_A-S_{A\not{B}})$ that appears in eq\ref{flow},\ref{discreteT} is then the difference of these two entanglement measures, in units of ebits. $\mathbb{T}_{B\rightarrow A}$ then quantifies the causal influence of B on A in the sense of how much it causes the entanglement of A with the rest of the universe to change. Similarly, other interpretations of von-Neumann entropy, such as uncertainty of a given state, also applies here. 

\paragraph*{Evolution of subsystem A with B frozen}
Since $\varepsilon(t)_{A\not{B}}$ is a mapping of density states, it can be interpreted as a quantum channel acting on subsystem A\cite{nielsen2002quantum}: $\rho_A(0)\overset{\varepsilon(t)_{A\not{B} }}{\rightarrow}\rho_{A\not{B}}(t)$. We further require that $\varepsilon(t)_{A\not{B}}$ corresponds to a physical process, therefore it can be obtained from taking the partial trace of the full system, which evolves unitarily. For tripartite system $\rho_{ABC}$:
\begin{equation}
\rho_{A\not{B}}(t)=\mathrm{Tr}_{BC}\{U_{A\not{B}C}(t)\rho_{ABC}(0 )U^\dagger_{A\not{B}C}(t)\}
\end{equation}
for some unitary operator $U_{A\not{B}C}$. 

Moreover, we require that the evolution mechanism with some subsystems frozen takes product form between the frozen qubits and the rest of the system: 
\begin{equation}
U_{A\not{B}C}(t)=\mathcal{V}_{AC}\otimes \mathcal{W}_B \label{UnotB}
\end{equation}
where $\mathcal{V}_{AC}$ and $\mathcal{W}_B$ are unitary operators acting on subsystems AC and B respectively. Frozen mechanism of the form Eq\ref{UnotB} guarantees what Liang referred to as \textit{the principle of nil causality}\cite{san2016information} (See SM for proof): 

\vspace{0.2cm}
\emph{$T_{B\rightarrow A}=0$ if the evolution of A is independent of B, that is, the unitary evolution operator $U_{ABC}(t)$ takes separable form $\mathcal{M}_A \otimes \mathcal{N}_{BC}$ or $\mathcal{O}_{AC} \otimes \mathcal{Q}_{B}$. }  
\vspace{0.2cm}

Therefore, the causal structure of space-time in physics is embedded in the formalism. If quantum operations, conducted at 4-dimensional coordinate $x$ and $y$, are space-like separated, hence non-causal, then the operations acting at $x$ does not affect the state located at $y$ and vice versa. The quantum operations at $x$ and $y$ commute and the joint evolution is in product form. Thus the quantum Liang information flow from one coordinate to another vanishes.

\subparagraph*{Bipartite system}
Consider bipartite state $\rho_{AB}$ under unitary evolution $U_{AB}(t)$. Consulting with eq\ref{UnotB}, $U_{A\not{B}}$ takes the form $\mathcal{V}_A\otimes \mathcal{W}_B$ in 2 dimensions. Since von-Neumann entropy is invariant under unitary change of basis, $\rho_{A\not{B}}=\mathcal{V}_A \rho_A(0) \mathcal{V}_A^\dagger$ and $\frac{dS_{A\not{B}}}{dt}=0$. Therefore, the rate of information flow from B to A: $T_{B\rightarrow A}=\frac{dS_A}{dt}$. Similarly,$T_{A\rightarrow B}=\frac{dS_B}{dt}$. If the initial state $\rho_{AB}(0)$ is pure, that is, the system is closed, by Schmidt decomposition, $\rho_A$ and $\rho_B$ share the same set of eigenvalues. Since closed bipartite system is symmetric, $S_A(t)=S_B(t)$ and $T_{B\rightarrow A}=T_{A\rightarrow B}$. In general, if the initial state $\rho_{AB}(0)$ is mixed, which can arise from entanglement with some external system, then we no longer have the symmetry $T_{A\rightarrow B}\neq T_{B\rightarrow A}$. Consider CNOT gate with controlled qubit A and target qubit B acts on the initial state $\rho_{AB}(0)=(1/2|0\rangle \langle 0|_A+1/2|1\rangle \langle 1|_A)\otimes|0\rangle \langle 0|_B$, the system evolves to $1/2|0\rangle \langle 0|_A\otimes|0\rangle \langle 0|_B+1/2|1\rangle \langle 1|_A\otimes|1\rangle \langle 1|_B$. The cumulative information flow for this discrete mapping $\mathbb{T}_{B\rightarrow A}=\Delta S_A=0$ and  $\mathbb{T}_{A\rightarrow B}=\Delta S_B=1bit$. The asymmetric quantum information flow obtained for initially mixed bipartite system parallels its classical counterpart (see SM for details).  For multi-partite system $\rho_{ABCD\cdots}$, the information flow from the rest of a closed system towards a particular unit, say A, is equivalent to the bipartite scenario: $T_{BCD\cdots \rightarrow A}=\frac{dS_A}{dt}$, $\mathbb{T}_{BCD\cdots \rightarrow A}=\Delta S_A$.
\subparagraph{Multipartite system}
In general, evaluation of the information flow in multipartite system requires a method to fix $\mathcal{V}_{AC}$ in eq\Ref{UnotB}. In this section, we illustrate such an approach with tripartite system $\rho_{ABC}$. Consider time evolution operator $U(t)=e^{-iHt/\hbar}$. We define the evolution of A with B frozen by replacing the interaction terms relevant to B in the Hamiltonian $H$, with identity operator. For instance, take Hamiltonian:
\begin{equation}
H_{ABC}=H_{0A}+H_{0B}+H_{0C}+\mathcal{A}\otimes \mathcal{C}+\mathcal{B} \otimes \mathcal{C} \label{Hamiltonian}
\end{equation} 
where $H_{0i}$,with $i=A,B,C$, is the free Hamiltonian. And $\mathcal{A},\mathcal{B},\mathcal{C}$, which describe their interactions, are hermitian operators acting on subsystem A, B, C respectively. The evolution mechanism with B frozen is then: $U_{A\not{B}C}=e^{-iH_{A\not{B}C} t/\hbar}$, where
\begin{equation}
H_{A\not{B}C}\equiv H_{0A}+H_{0C}+\mathcal{A}\otimes \mathcal{C}+I_B
\end{equation}
$U_{A\not{B}C}$ is clearly of the product form given in eq\ref{UnotB}, with $\mathcal{W}_B=I$ and $\mathcal{V}_{AC}$ generated by hermitian operator $H_{0A}+H_{0C}+\mathcal{A}\otimes \mathcal{C}$. The operational meaning of $U_{A\not{B}C}$ is then:

\vspace{0.2cm}
\emph{evolution of the system if subsystem B is removed from the original evolution mechanism.}  
\vspace{0.2cm}

\begin{figure}[h!]
	\centering
	\subfigure[]{%
		\includegraphics[width=0.2\textwidth]{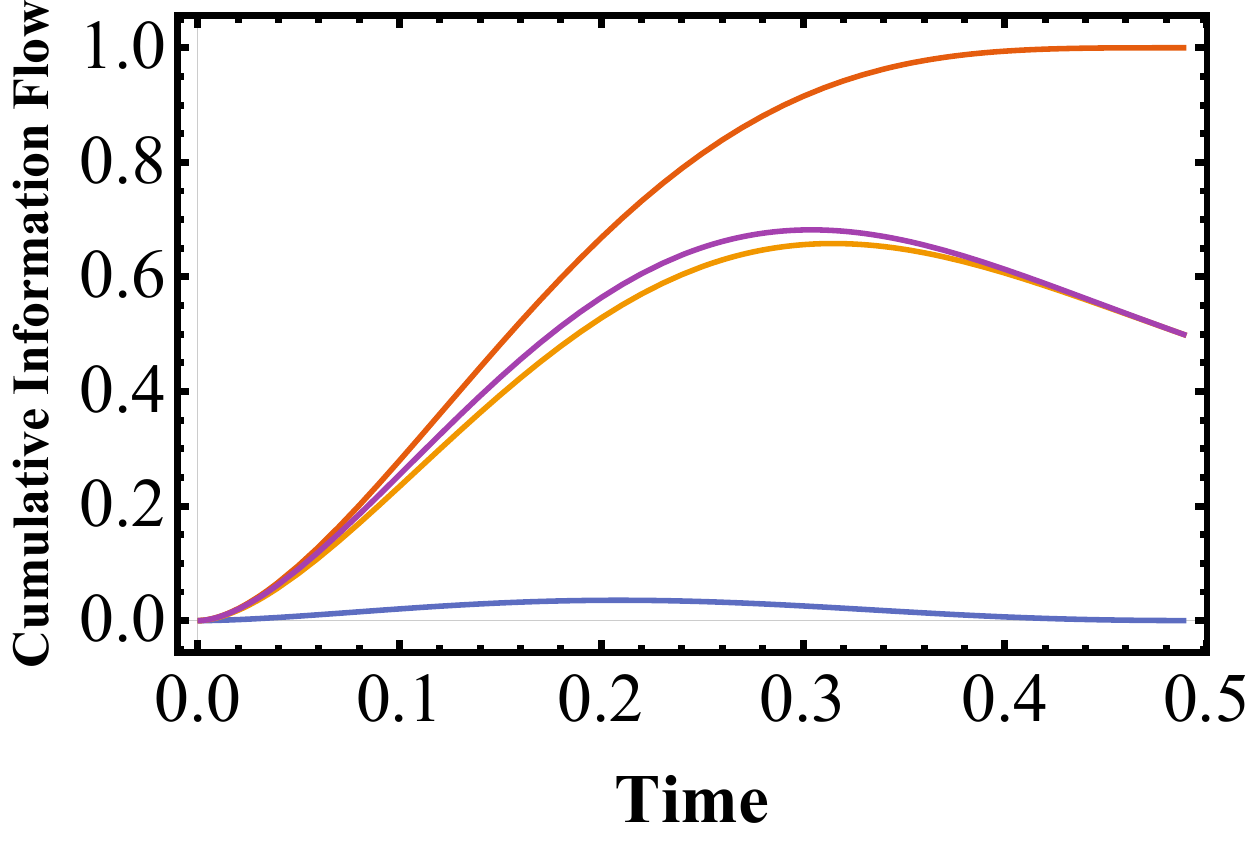}%
		\label{3spin1}%
	}\hfil
	\subfigure[]{%
		\includegraphics[width=0.2\textwidth]{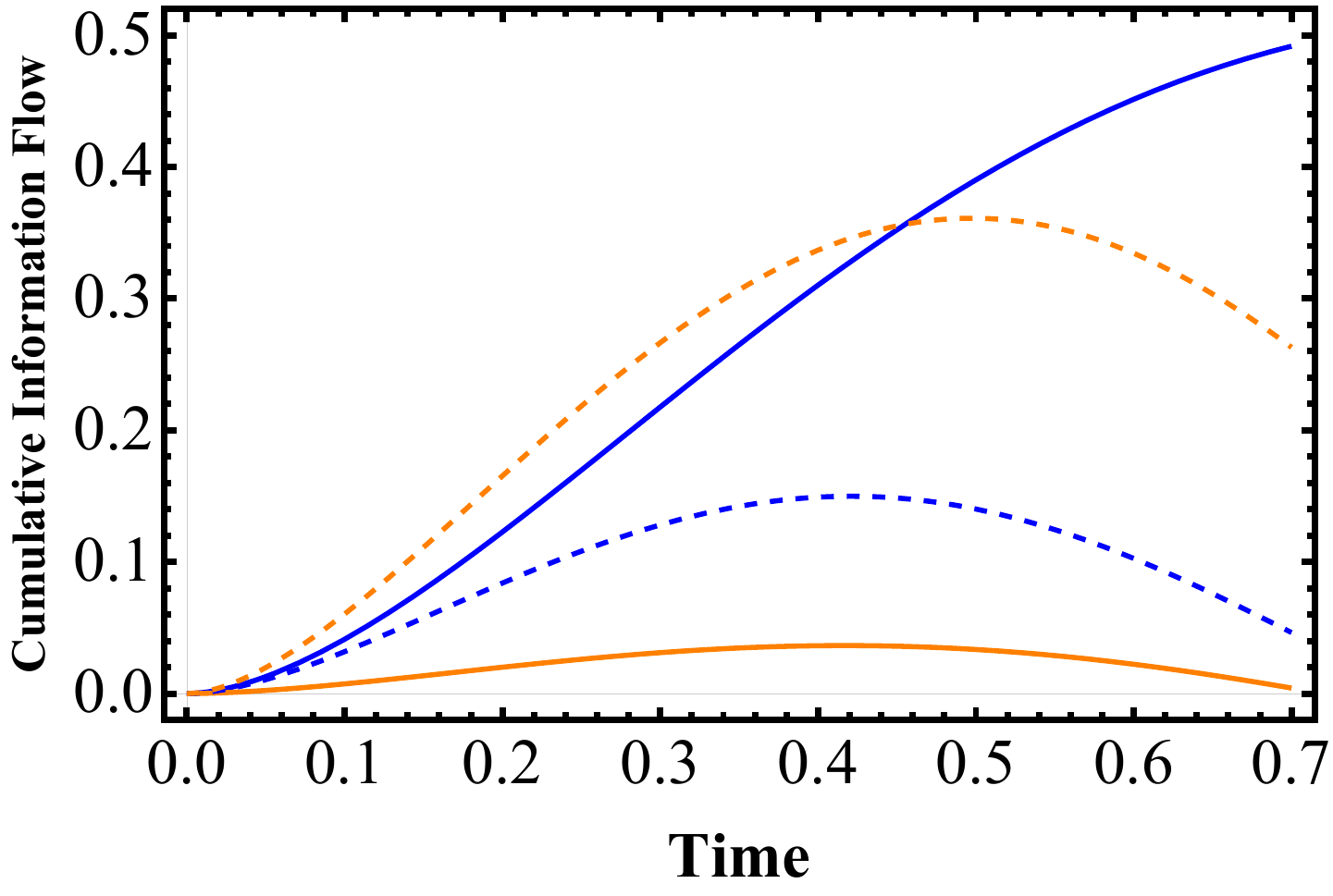}%
		\label{3spininitial}%
	}
	
	\caption{\footnotesize \textbf{3-qubit spin chain} (a) From top to bottom (measured in Bits): $\mathbb{T}_{AB\rightarrow C}$, $\mathbb{T}_{B\rightarrow C}+\mathbb{T}_{A\rightarrow C}$, $\mathbb{T}_{B\rightarrow C}$, $\mathbb{T}_{A\rightarrow C}$. Coupling strength: $\eta_{AC}=1$, $\eta_{BC}=3$. Initial state: $\rho(0)=I_A\otimes I_B\otimes |0\rangle \langle 0|_C$. (b)Blue curves: $\mathbb{T}_{A\rightarrow C}$, Orange curves: $\mathbb{T}_{B\rightarrow C}$. Solid curves: Initial state $\rho_{0(1)}=I_A\otimes (0.9|0\rangle \langle 0|+0.1|1\rangle \langle 1|)_{B}\otimes|0\rangle \langle 0|_C$, Dashed curves: Initial state $\rho_{0(2)}=I_A\otimes (0.1|0\rangle \langle 0|+0.9|1\rangle \langle 1|)_{B}\otimes|0\rangle \langle 0|_C$. Coupling strength: $\eta_{AC}=\eta_{BC}=1$.}
\end{figure}

The operational meaning of the frozen mechanism guarantees that this definition is basis(observable) independent. Now, we are equipped with the tools needed to evaluate quantum Liang information flow. In the next section, we will elucidate this formalism with applications. 
\paragraph*{Application: multi-qubit spin system}
Consider a multi-qubit spin chain, the interaction Hamiltonian between any two interacting qubits i,j is given by\cite{yung2003exact}:
\begin{equation}
H_{spin,ij}=\eta_{ij} (\sigma_{+i}\sigma_{-j}+\sigma_{-i}\sigma_{+j}) \label{spinH}
\end{equation}
where $\sigma_{\pm}$ can be expressed in terms of Pauli matrices  $\{\sigma_{x,y,z}\}$, $\sigma_{\pm}=\frac{1}{2}(\sigma_x\pm i \sigma_y)$. $\eta$ is relative coupling strength. The Hamiltonian for 3 interacting qubits, labeled A, B, C, of the form eq\ref{Hamiltonian} is given by:
\begin{equation}
\eta_{AC} (\sigma_{+A}\sigma_{-C}+\sigma_{-A}\sigma_{+C })+\eta_{BC}(\sigma_{+B}\sigma_{-C}+\sigma_{-B}\sigma_{+C}) \label{H3spin}
\end{equation}
\subparagraph*{Relative coupling strength variation}
In this section, we investigate cumulative Information flow $\mathbb{T}$ from A, B to C with different coupling strength. We set the initial state of the sending qubits A, B being maximally mixed while the receiving qubits C pure: $\rho(0)=I_A\otimes I_{B}\otimes|0\rangle\langle 0|_C$. So the sending qubits are competing to propagate uncertainty towards the target qubit. The Hamiltonian with one qubit frozen, say A, is obtained by erasing the terms involving qubit A in Hamiltonian eq\ref{H3spin}:
\begin{equation}
H_{\not{A}BC}=\lambda\sigma_{+C}\sigma_{-B}+\sigma_{-C}\sigma_{+B}+I_A \label{spinAnotB}
\end{equation}
The evolution of $\rho_{\not{A}\not{B}C}$ is defined similarly by removing hermitian terms relevant to qubits A,B altogether. Therefore, $\Delta S_{\not{A}\not{B}C}$ vanishes and the joint cumulative information flow from AB to C is: $\mathbb{T}_{AB\rightarrow C}=\Delta S_C$. Set $\eta_{AC}=1$, $\eta_{BC}=3$, at time $t\sim 0.49$, the entropy of C reaches its maxima of 1 bit for the first time. This is the maximum uncertainty qubit C can receive, determined by its dimension. For the purpose of illustration, we compare the cumulative information flow from different sending qubits before this capacity is reached. The early time behavior of cumulative information flow $\mathbb{T}_{AB\rightarrow C}(t)$, $\mathbb{T}_{A\rightarrow C}(t)$, $\mathbb{T}_{B\rightarrow C}(t)$ is plotted in figure \ref{3spin1}.

From figure\ref{3spin1}, we notice that: The cumulative information flow from B to C is greater than that from A to C: $\mathbb{T}_{B\rightarrow C}>\mathbb{T}_{A\rightarrow C}$. This formalism is consistent with the intuition that the strongly coupled qubit has greater impact towards the target. The direct addition of cumulative information flow from individual qubit A, B is smaller than the joint value: $\mathbb{T}_{B\rightarrow C}+\mathbb{T}_{A\rightarrow C}<\mathbb{T}_{AB\rightarrow C}$ in this example. It means that turning off qubit A and B altogether has more impact on qubit C than the direct addition of turning A, B off one at a time. Similar result is obtained for the early time behavior of 5 qubit spin chain (See SM).
\subparagraph*{Initial configuration dependence}
Note that the information flow formalism also depends on the initial configuration. To see how different initial states affect the information flow, set the coupling constant equal: $\eta_{AC}=\eta_{BC}=1$, with initial state $\rho_{0(1)}=I_A\otimes (0.9|0\rangle \langle 0|+0.1|1\rangle \langle 1|)_{B}\otimes|0\rangle \langle 0|_C$ and $\rho_{0(2)}=I_A\otimes (0.1|0\rangle \langle 0|+0.9|1\rangle \langle 1|)_{B}\otimes|0\rangle \langle 0|_C$. In both cases, the initial entropy of qubit B is $\sim0.47$bit while A is 1 bit. At a first glance, one may be expecting that A is transmitting more uncertainty to C than qubit B. From figure\ref{3spininitial}, we see this is indeed the case for initial state $\rho_{0(1)}$. But when the initial state is switched to $\rho_{0(2)}$, we have $\mathbb{T}_{B\rightarrow C}>\mathbb{T}_{A\rightarrow C}$. This is because increasing in von-Neumann entropy could result from not only classical uncertainty propagation but also entanglement generation. The qubit interaction given in eq\ref{spinH} entangles state $|10\rangle$ ($|01\rangle$), while it does not act on state $|00\rangle$($|11\rangle$):
\begin{eqnarray}
&(\sigma_{+}\sigma_{-}+\sigma_{-}\sigma_{+})|00\rangle=0\nonumber ,\,\,\ (\sigma_{+}\sigma_{-}+\sigma_{-}\sigma_{+})|10\rangle=|01\rangle \nonumber
\end{eqnarray}
For initial state $\rho_{0(2)}$, qubit B and C has 90\% probability in $|10\rangle_{BC}$ state, the entangling mechanism greatly increases $\mathbb{T}_{B\rightarrow C}$ compare to $\rho_{0(1)}$, for which the probability is only 10\%. Changing the initial state to $\rho_{0(2)}$ also suppresses $\mathbb{T}_{A\rightarrow C}$ due to growing competition from B.

\begin{figure}[h!]
	\centering
	\subfigure[]{%
		\includegraphics[width=0.2\textwidth]{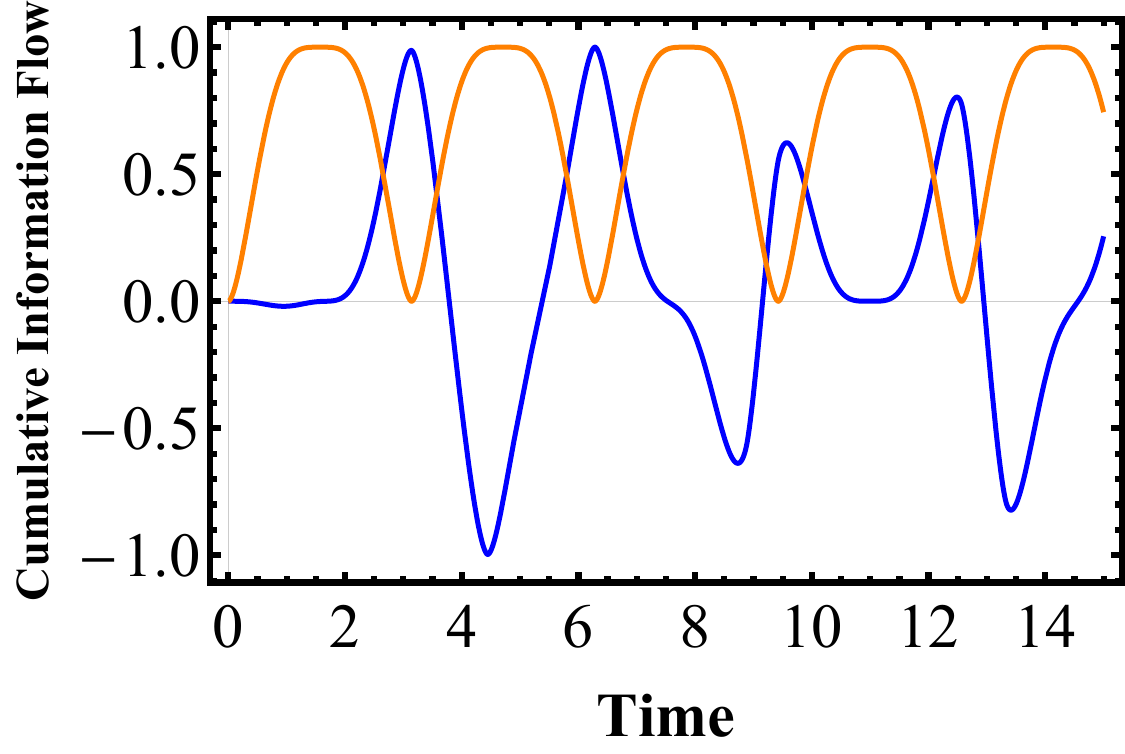}%
		\label{superexchange1}%
	}\hfil
		\subfigure[]{%
		\includegraphics[width=0.2\textwidth]{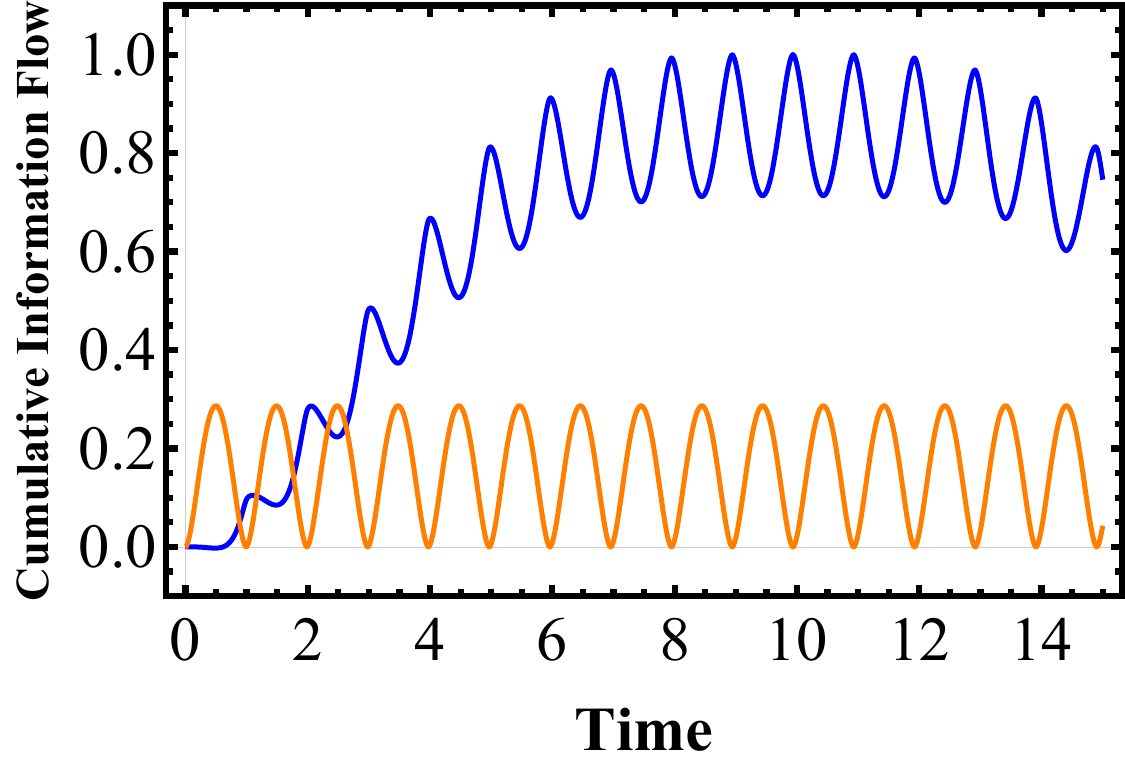}%
		\label{superexchange2}%
	}
	
	\subfigure[]{%
		\includegraphics[width=0.2\textwidth]{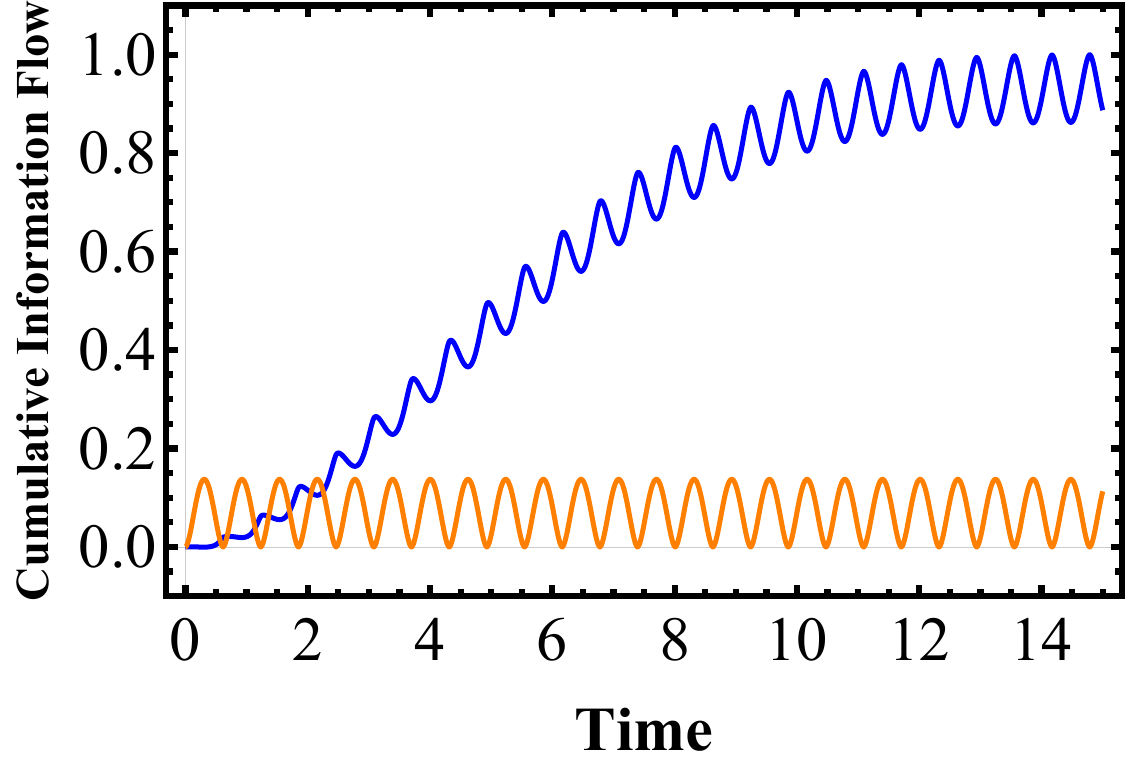}%
		\label{superexchange3}%
	}\hfil
	\subfigure[]{%
		\includegraphics[width=0.2\textwidth]{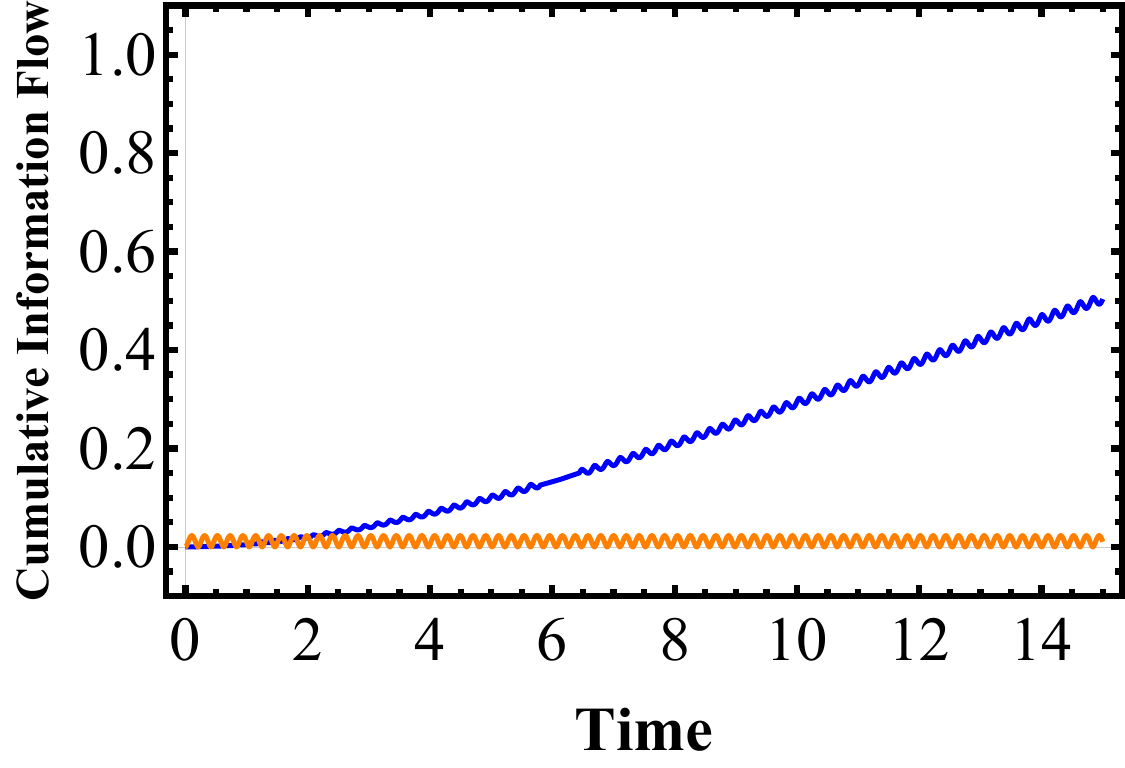}%
		\label{superexchange4}%
	}
	
	\caption{\footnotesize \textbf{Quantum super-exchange}: (In Bits) Blue curves: $\mathbb{T}_{A\rightarrow B}$, Orange curves: $\mathbb{T}_{C\rightarrow B}$. In \ref{superexchange1},\ref{superexchange2},\ref{superexchange3},\ref{superexchange4}, Magnetic field strength set to $\mathbf{B}=0, 3, 5, 15$ respectively. Coupling strength: $\eta_{AC}=\eta_{BC}=1$. Initial state: $I_A\otimes|0\rangle \langle 0|_B\otimes I_{C}$. }\label{superexchange}
\end{figure}

\subparagraph*{Quantum super-exchange}
 Add constant magnetic field along the z-axis with strength $\mathbf{B}$ on the intermediate qubit C so that its energy is lifted by an amount $\mathbf{B}\sigma_z$, while qubit A and B remains unaffected. The total Hamiltonian acting on the system then adds up an additional term:
 \begin{equation}
H_{additional}=I_A\otimes  I_B \otimes \mathbf{B}\sigma_{z(C)} 
 \end{equation}
 Set coupling strength $\eta_{AC}=\eta_{BC}=1$ and initial state $\rho(0)=I_A\otimes|0\rangle \langle 0|_B\otimes I_{C}$. We wish to compare information flow from A,C to B with various magnetic field strength. Note that when $\mathbf{B}=0$, the dynamics of information flow in the XY model (eq\ref{spinH}), which is not apriori obvious, can be pictured from fig\ref{superexchange1} The cumulative information flow is initially from C to B and it reaches a high value of 1 bit before it declines and is overtaken by the cumulative information flow from A to B. As the magnetic field strength increases, super-exchange process \cite{benjamin2004quantum} between A and B becomes progressively dominant. Thus, we see that information flow from C to B goes down while that from A to B becomes that dictated by an effective weaker super-exchange coupling $\eta_{AC}^2/\mathbf{B}$ between A and B $(\sigma_{+A}\sigma_{-B}+h.c.)$\cite{benjamin2004quantum}.

\begin{figure}[h!]
	\centering
	\subfigure[]{%
		\includegraphics[width=0.2\textwidth]{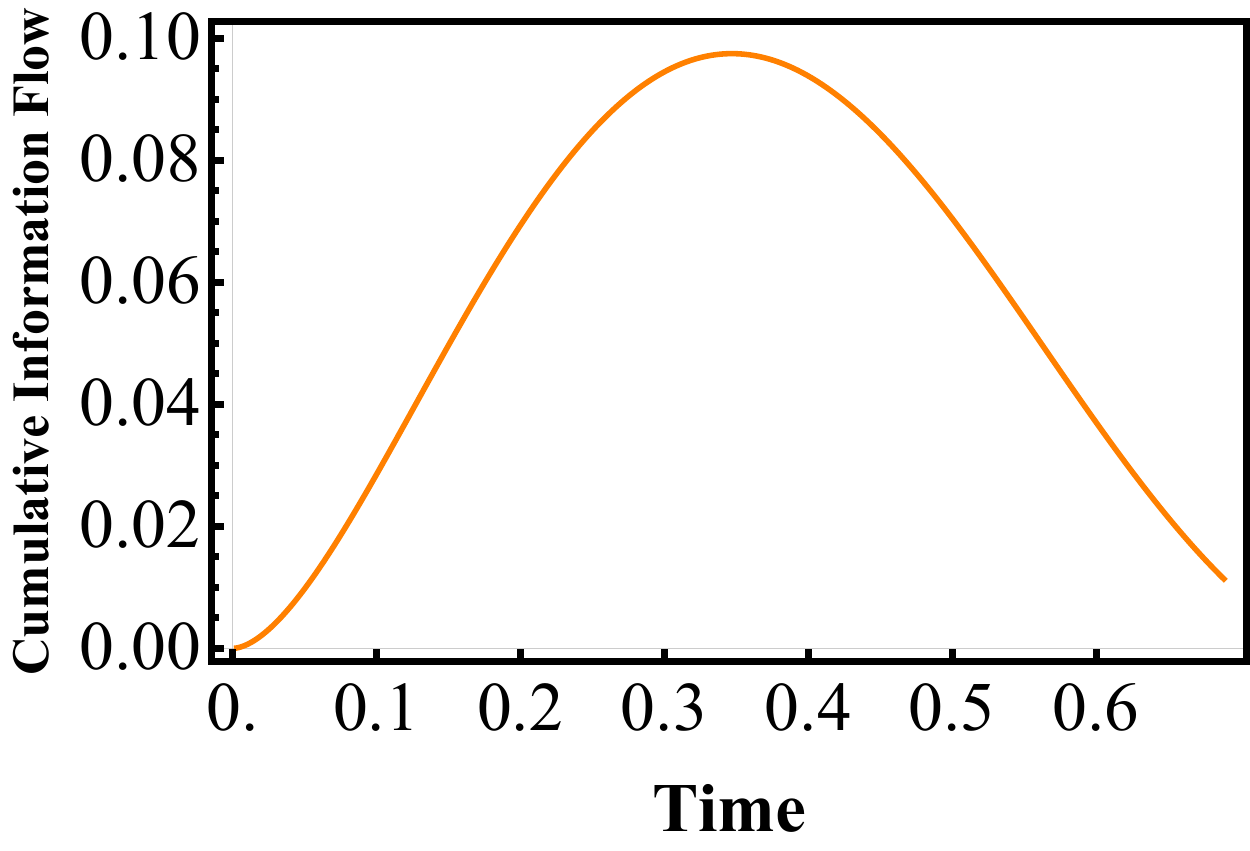}%
		\label{5qubit1}%
	}\hfil
	\subfigure[]{%
		\includegraphics[width=0.2\textwidth]{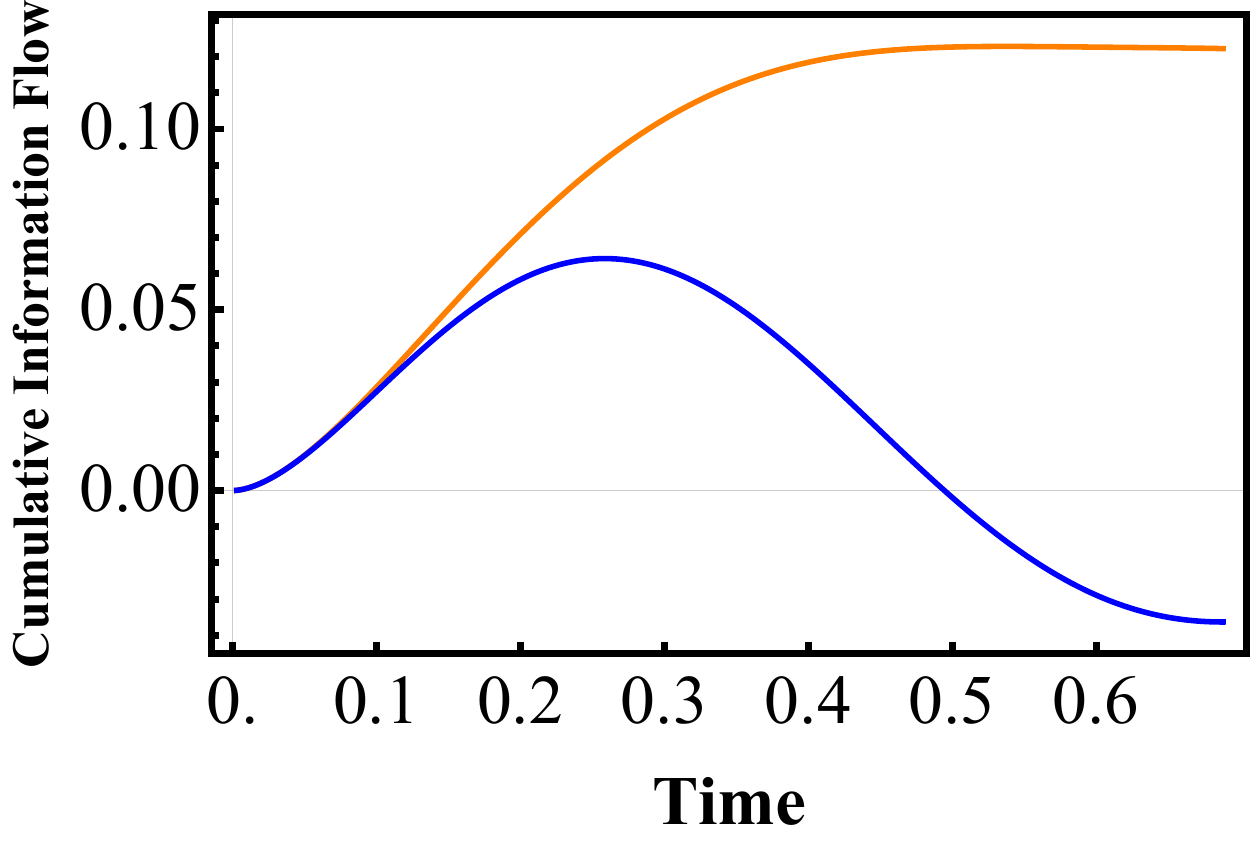}%
		\label{5qubit2}%
	}
	
	\caption{\footnotesize \textbf{5-qubit network} Cumulative information flow (in Bits) (a) from any sending qubit towards E with identical coupling strength: $\eta_{DE}=\eta_{CE}=\eta_{BE}=\eta_{AE}=1$. (b) with additional coupling $\eta_{CD}=5$. Orange curve: A(or B) to E, Blue curve: C(or D) to E.}
\end{figure}

\subparagraph*{5-qubit network}
Consider 5-qubit spin system, labeled A,B,C,D,E, with E in the center, we wish to investigate information flow towards E. The total Hamiltonian for the 5-qubit spin chain is
\begin{equation}
H_{spin,tot}=\sum_{i} H_{spin,iE}
\end{equation} 
with $i=A,B,C,D$. Set all the coupling strength with E identical: $\eta_{DE}=\eta_{CE}=\eta_{BE}=\eta_{AE}=1$, and initial state of sending qubits A,B,C,D maximally mixed, receiving qubit E pure. At time $t\sim0.69$, the entropy of E reaches its maximum of 1 bit for the first time. The cumulative information flow from each sending qubit, which is identical $\mathbb{T}_{A\rightarrow E}=\mathbb{T}_{B\rightarrow E}=\mathbb{T}_{C\rightarrow E}=\mathbb{T}_{D \rightarrow E}$, is plotted for the time interval $t\in [0,0.69]$ in figure\ref{5qubit1}. 

Now let us add mutual interaction between C, D with relative coupling strength $\eta_{CD}=5$ and observe how does the information flow towards the center qubit E changes (schematic diagram of the interaction pattern is shown in SM). The total Hamiltonian is now given by:
\begin{equation}
\sum_{i} H_{spin,iE}+H_{spin,CD}
\end{equation}

With presence of this additional interaction term, the cumulative information flow from each sending qubit to E is plotted in figure\ref{5qubit2}. Compare figure\ref{5qubit2} with figure\ref{5qubit1}, the presence of the additional interaction term between C,D greatly reduces the transmitted uncertainty from qubit C (D) to qubit E, while increases that from qubit A (B) to qubit E. After time $t\sim0.49$, $\mathbb{T}_{C\rightarrow E}$ reaches negative value, that is, the presence of qubit C (D) reduces the uncertainty of qubit E. The uncertainty from qubit C (D) now has two routes to propagate, either towards E or D (C). Also, the relative coupling strength $\eta_{CD}$ is five times stronger than $\eta_{CE},\eta_{DE}$. The strongly coupled route connecting C and D then diverts the uncertainty propagation away from the original path between C (D) and E, so that $\mathbb{T}_{C\rightarrow E}$($\mathbb{T}_{D\rightarrow E}$) decreases. Qubit A and B now has less competition from qubit C and D to propagate uncertainty towards qubit E. Then,  $\mathbb{T}_{A\rightarrow E}$($\mathbb{T}_{B\rightarrow E}$) increases. The presence of certain coupling could preserve information, which may be exploited to design robust quantum circuits.

\paragraph*{Application: Two-qubit system in bosonic bath}
Let subsystem A and B indicate two non-interacting qubits with ground and excited states $|0\rangle$, $|1\rangle$, embedded in a common zero-temperature bosonic reservoir labeled C. We wish to compare information flow between the two qubits. The Hamiltonian governing the mechanism is given by $H_{SB}=H_0+H_{int}$, with:
\begin{eqnarray}
H_0&=&\omega_0 \sigma_+^A \sigma_-^A+ \omega_0 \sigma_+^B \sigma_-^B+\sum_{k}\omega_k b_k^\dagger b_k \nonumber\\
H_{int}&=&\alpha_A \sigma_+^A \sum_{k}g_k b_k+\alpha_B \sigma_+^B \sum_{k}g_k b_k+h.c. \label{Hsb}
\end{eqnarray}
where $\sigma_{\pm}^{A(B)}$ and $\omega_0$ are the inversion operator and transition frequency of qubit A(B). $b_k,b_k^\dagger$ are annihilation and creation operator of the environment C.  $\alpha_{A(B)}$ measures the coupling between each qubit and the reservoir. In the limit $\alpha_B$ or $\alpha_A$ goes to 0, that is, when one of the qubit decouples from the setup, then $\rho_A$ and $\rho_{A\not{B}}$ obeys the same equation of motion and $\rho_A(t)=\rho_{A\not{B}}(t)$. Therefore, $T_{B\rightarrow A}=T_{A\rightarrow B}=0$.

\begin{figure}[h!]
	\centering
	\subfigure[]{%
		\includegraphics[width=0.2\textwidth]{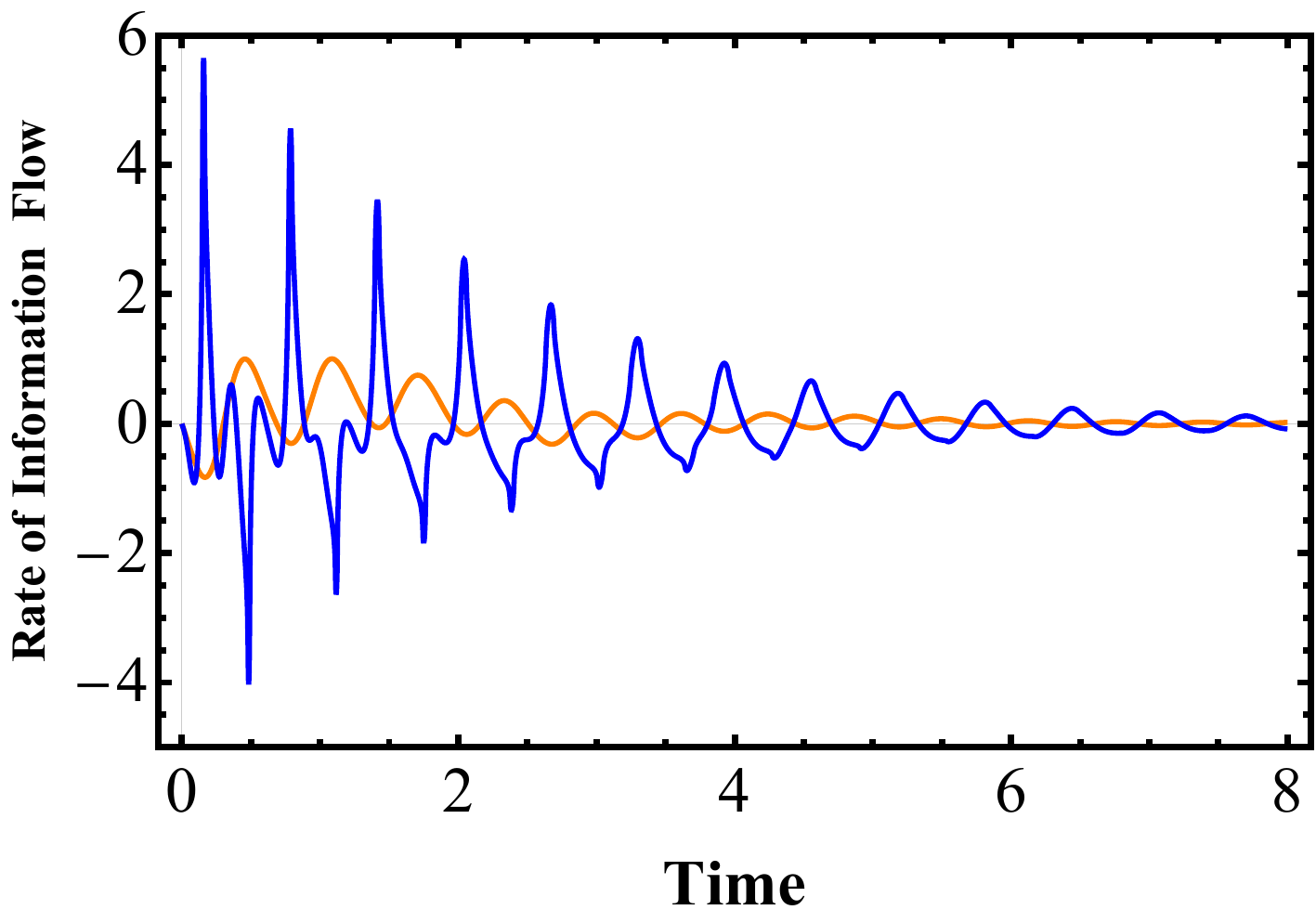}%
		\label{spinboson1}%
	}\hfil
	\subfigure[]{%
		\includegraphics[width=0.2\textwidth]{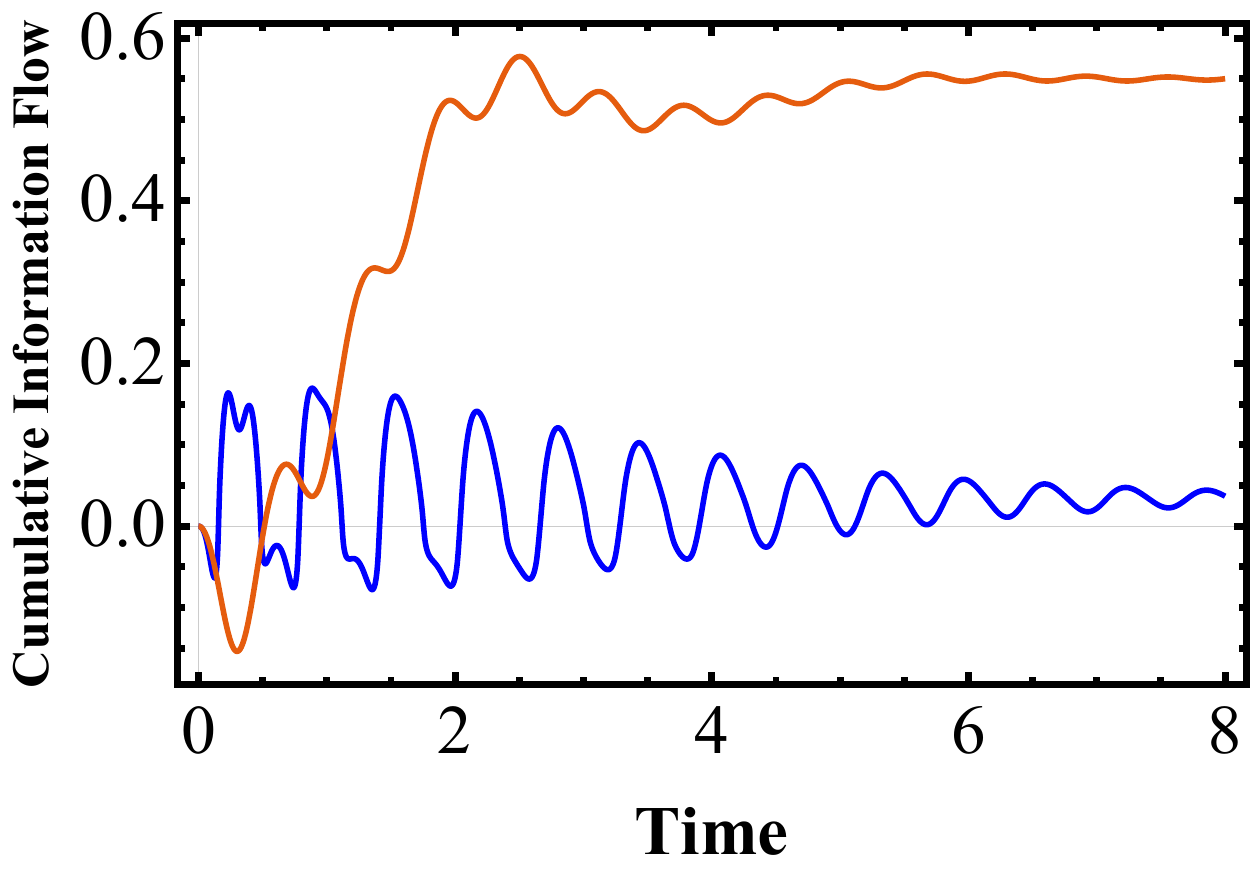}%
		\label{spinboson2}%
	}
	
	\caption{\footnotesize \textbf{Two-qubit system in a lossy cavity} Blue curves: From B to A. Orange curves: From A to B. Coupling strength ratio: $\alpha_A/\alpha_B=10/1$. (a) Rate of information flow (Bits per unit time) (b) Cumulative information flow (Bits).}
\end{figure}

 We adopt the lossy cavity model given in Ref\cite{franco2013dynamics}. The two-qubit dynamics is solved exactly at zero temperature. Take initial state $\rho_{AB}(0)=|\psi_0\rangle \langle \psi_0|$, where $|\psi_0\rangle=\frac{1}{\sqrt{3}}(|01\rangle +\sqrt{2}|10\rangle)$. Let $\lambda=1$, $\hbar=1$, $\alpha_A/\alpha_B=10/1$ and take strong coupling limit $R=10$, where $\lambda$ defines the spectral width of the coupling and $R$ determines the collective coupling strength. The rate of information flow from B to A versus that from A to B is plotted in figure \ref{spinboson1}. The cumulative information flow is shown in figure \ref{spinboson2}. 
From Fig\ref{spinboson1}, we see that the rate of information flow from the weakly coupled qubit (B) towards the strongly coupled qubit (A) possess higher peak than that from A to B. On the other hand, as shown in fig\ref{spinboson2}, the cumulative information flow from A to B grows steadily and surpass that from B to A as the system approaches equilibrium. Note that the information flow formalism is generically asymmetric $T_{B\rightarrow A}\neq T_{A\rightarrow B}$ as opposed to most quantum correlation measures.

\paragraph*{Conclusions:}

In this paper, we have generalized Liang's methodology to quantify the causal influences in a quantum network. A unique feature of quantum networks is the possiblity of entanglement between its components. Thus, there are two ways to increase the entropy of a node: classical uncertainty propagation, as well as the growth of entanglement. We have verified the formalism through simple networks. The information flow between two qubits through a common bath could be nontrivial in the sense that the weakly coupled qubit has higher rate of information flow, while in the long run, the strongly coupled qubit has more impact on the weakly coupled one. Another non-trivial result obtained for a 5-qubit network reveals that an additional strong coupling diverts the directions of uncertainty propagation. Causal influences in generic complex quantum networks may be intricate and a picturization in terms of information flow will certainly aid their understanding. Note that definition of the information flow formalism is based on 1.full knowledge of the dynamics, 2.an intervention (frozen mechanism) act upon the system. For its classical counterpart, Liang has showed that the information flow can be estimated with local statistics for a broad range of subjects\cite{san2014unraveling,san2016information,liang2021measuring,liang2021normalized,vannitsem2019testing,hristopulos2019disrupted,hagan2019time,stips2016causal}. To what extent can the quantum version be estimated without knowing the dynamics apriori or doing the intervention on the system remains a subject for further investigation. 

\acknowledgements B.Yi would like to thank S.X.Huang for presenting the problem as well as inspiring discussions with X.S.Liang and A.J.Leggett. S.Bose acknowledges the EPSRC grant Nonergodic quantum manipulation EP/R029075/1.


\begin{thebibliography}{58}%
\makeatletter
\providecommand \@ifxundefined [1]{%
 \@ifx{#1\undefined}
}%
\providecommand \@ifnum [1]{%
 \ifnum #1\expandafter \@firstoftwo
 \else \expandafter \@secondoftwo
 \fi
}%
\providecommand \@ifx [1]{%
 \ifx #1\expandafter \@firstoftwo
 \else \expandafter \@secondoftwo
 \fi
}%
\providecommand \natexlab [1]{#1}%
\providecommand \enquote  [1]{``#1''}%
\providecommand \bibnamefont  [1]{#1}%
\providecommand \bibfnamefont [1]{#1}%
\providecommand \citenamefont [1]{#1}%
\providecommand \href@noop [0]{\@secondoftwo}%
\providecommand \href [0]{\begingroup \@sanitize@url \@href}%
\providecommand \@href[1]{\@@startlink{#1}\@@href}%
\providecommand \@@href[1]{\endgroup#1\@@endlink}%
\providecommand \@sanitize@url [0]{\catcode `\\12\catcode `\$12\catcode
  `\&12\catcode `\#12\catcode `\^12\catcode `\_12\catcode `\%12\relax}%
\providecommand \@@startlink[1]{}%
\providecommand \@@endlink[0]{}%
\providecommand \url  [0]{\begingroup\@sanitize@url \@url }%
\providecommand \@url [1]{\endgroup\@href {#1}{\urlprefix }}%
\providecommand \urlprefix  [0]{URL }%
\providecommand \Eprint [0]{\href }%
\providecommand \doibase [0]{http://dx.doi.org/}%
\providecommand \selectlanguage [0]{\@gobble}%
\providecommand \bibinfo  [0]{\@secondoftwo}%
\providecommand \bibfield  [0]{\@secondoftwo}%
\providecommand \translation [1]{[#1]}%
\providecommand \BibitemOpen [0]{}%
\providecommand \bibitemStop [0]{}%
\providecommand \bibitemNoStop [0]{.\EOS\space}%
\providecommand \EOS [0]{\spacefactor3000\relax}%
\providecommand \BibitemShut  [1]{\csname bibitem#1\endcsname}%
\let\auto@bib@innerbib\@empty
\bibitem [{\citenamefont {Hlav{\'a}{\v{c}}kov{\'a}-Schindler}\ \emph
  {et~al.}(2007)\citenamefont {Hlav{\'a}{\v{c}}kov{\'a}-Schindler},
  \citenamefont {Palu{\v{s}}}, \citenamefont {Vejmelka},\ and\ \citenamefont
  {Bhattacharya}}]{hlavavckova2007causality}%
  \BibitemOpen
  \bibfield  {author} {\bibinfo {author} {\bibfnamefont {K.}~\bibnamefont
  {Hlav{\'a}{\v{c}}kov{\'a}-Schindler}}, \bibinfo {author} {\bibfnamefont
  {M.}~\bibnamefont {Palu{\v{s}}}}, \bibinfo {author} {\bibfnamefont
  {M.}~\bibnamefont {Vejmelka}}, \ and\ \bibinfo {author} {\bibfnamefont
  {J.}~\bibnamefont {Bhattacharya}},\ }\href@noop {} {\bibfield  {journal}
  {\bibinfo  {journal} {Physics Reports}\ }\textbf {\bibinfo {volume} {441}},\
  \bibinfo {pages} {1} (\bibinfo {year} {2007})}\BibitemShut {NoStop}%
\bibitem [{\citenamefont {Pearl}\ \emph {et~al.}(2000)\citenamefont {Pearl}
  \emph {et~al.}}]{pearl2000models}%
  \BibitemOpen
  \bibfield  {author} {\bibinfo {author} {\bibfnamefont {J.}~\bibnamefont
  {Pearl}} \emph {et~al.},\ }\href@noop {} {\bibfield  {journal} {\bibinfo
  {journal} {Cambridge, UK: CambridgeUniversityPress}\ }\textbf {\bibinfo
  {volume} {19}} (\bibinfo {year} {2000})}\BibitemShut {NoStop}%
\bibitem [{\citenamefont {Spirtes}\ \emph {et~al.}(2000)\citenamefont
  {Spirtes}, \citenamefont {Glymour}, \citenamefont {Scheines},\ and\
  \citenamefont {Heckerman}}]{spirtes2000causation}%
  \BibitemOpen
  \bibfield  {author} {\bibinfo {author} {\bibfnamefont {P.}~\bibnamefont
  {Spirtes}}, \bibinfo {author} {\bibfnamefont {C.~N.}\ \bibnamefont
  {Glymour}}, \bibinfo {author} {\bibfnamefont {R.}~\bibnamefont {Scheines}}, \
  and\ \bibinfo {author} {\bibfnamefont {D.}~\bibnamefont {Heckerman}},\
  }\href@noop {} {\emph {\bibinfo {title} {Causation, prediction, and
  search}}}\ (\bibinfo  {publisher} {MIT press},\ \bibinfo {year}
  {2000})\BibitemShut {NoStop}%
\bibitem [{\citenamefont {Bezruchko}\ and\ \citenamefont
  {Smirnov}(2010)}]{bezruchko2010extracting}%
  \BibitemOpen
  \bibfield  {author} {\bibinfo {author} {\bibfnamefont {B.~P.}\ \bibnamefont
  {Bezruchko}}\ and\ \bibinfo {author} {\bibfnamefont {D.~A.}\ \bibnamefont
  {Smirnov}},\ }\href@noop {} {\emph {\bibinfo {title} {Extracting knowledge
  from time series: An introduction to nonlinear empirical modeling}}}\
  (\bibinfo  {publisher} {Springer Science \& Business Media},\ \bibinfo {year}
  {2010})\BibitemShut {NoStop}%
\bibitem [{\citenamefont {Schreiber}(2000)}]{schreiber2000measuring}%
  \BibitemOpen
  \bibfield  {author} {\bibinfo {author} {\bibfnamefont {T.}~\bibnamefont
  {Schreiber}},\ }\href@noop {} {\bibfield  {journal} {\bibinfo  {journal}
  {Physical review letters}\ }\textbf {\bibinfo {volume} {85}},\ \bibinfo
  {pages} {461} (\bibinfo {year} {2000})}\BibitemShut {NoStop}%
\bibitem [{\citenamefont {Wang}\ \emph {et~al.}(2004)\citenamefont {Wang},
  \citenamefont {Anderson}, \citenamefont {Kaufmann},\ and\ \citenamefont
  {Myneni}}]{wang2004relation}%
  \BibitemOpen
  \bibfield  {author} {\bibinfo {author} {\bibfnamefont {W.}~\bibnamefont
  {Wang}}, \bibinfo {author} {\bibfnamefont {B.~T.}\ \bibnamefont {Anderson}},
  \bibinfo {author} {\bibfnamefont {R.~K.}\ \bibnamefont {Kaufmann}}, \ and\
  \bibinfo {author} {\bibfnamefont {R.~B.}\ \bibnamefont {Myneni}},\
  }\href@noop {} {\bibfield  {journal} {\bibinfo  {journal} {Journal of
  Climate}\ }\textbf {\bibinfo {volume} {17}},\ \bibinfo {pages} {4752}
  (\bibinfo {year} {2004})}\BibitemShut {NoStop}%
\bibitem [{\citenamefont {Runge}\ \emph {et~al.}(2012)\citenamefont {Runge},
  \citenamefont {Heitzig}, \citenamefont {Marwan},\ and\ \citenamefont
  {Kurths}}]{runge2012quantifying}%
  \BibitemOpen
  \bibfield  {author} {\bibinfo {author} {\bibfnamefont {J.}~\bibnamefont
  {Runge}}, \bibinfo {author} {\bibfnamefont {J.}~\bibnamefont {Heitzig}},
  \bibinfo {author} {\bibfnamefont {N.}~\bibnamefont {Marwan}}, \ and\ \bibinfo
  {author} {\bibfnamefont {J.}~\bibnamefont {Kurths}},\ }\href@noop {}
  {\bibfield  {journal} {\bibinfo  {journal} {Physical Review E}\ }\textbf
  {\bibinfo {volume} {86}},\ \bibinfo {pages} {061121} (\bibinfo {year}
  {2012})}\BibitemShut {NoStop}%
\bibitem [{\citenamefont {Sun}(1994)}]{sun1994neural}%
  \BibitemOpen
  \bibfield  {author} {\bibinfo {author} {\bibfnamefont {R.}~\bibnamefont
  {Sun}},\ }\href@noop {} {\bibfield  {journal} {\bibinfo  {journal} {IEEE
  Transactions on Neural Networks}\ }\textbf {\bibinfo {volume} {5}},\ \bibinfo
  {pages} {604} (\bibinfo {year} {1994})}\BibitemShut {NoStop}%
\bibitem [{\citenamefont {Ay}\ and\ \citenamefont
  {Polani}(2008)}]{ay2008information}%
  \BibitemOpen
  \bibfield  {author} {\bibinfo {author} {\bibfnamefont {N.}~\bibnamefont
  {Ay}}\ and\ \bibinfo {author} {\bibfnamefont {D.}~\bibnamefont {Polani}},\
  }\href@noop {} {\bibfield  {journal} {\bibinfo  {journal} {Advances in
  complex systems}\ }\textbf {\bibinfo {volume} {11}},\ \bibinfo {pages} {17}
  (\bibinfo {year} {2008})}\BibitemShut {NoStop}%
\bibitem [{\citenamefont {Sommerlade}\ \emph {et~al.}(2009)\citenamefont
  {Sommerlade}, \citenamefont {Eichler}, \citenamefont {Jachan}, \citenamefont
  {Henschel}, \citenamefont {Timmer},\ and\ \citenamefont
  {Schelter}}]{sommerlade2009estimating}%
  \BibitemOpen
  \bibfield  {author} {\bibinfo {author} {\bibfnamefont {L.}~\bibnamefont
  {Sommerlade}}, \bibinfo {author} {\bibfnamefont {M.}~\bibnamefont {Eichler}},
  \bibinfo {author} {\bibfnamefont {M.}~\bibnamefont {Jachan}}, \bibinfo
  {author} {\bibfnamefont {K.}~\bibnamefont {Henschel}}, \bibinfo {author}
  {\bibfnamefont {J.}~\bibnamefont {Timmer}}, \ and\ \bibinfo {author}
  {\bibfnamefont {B.}~\bibnamefont {Schelter}},\ }\href@noop {} {\bibfield
  {journal} {\bibinfo  {journal} {Physical Review E}\ }\textbf {\bibinfo
  {volume} {80}},\ \bibinfo {pages} {051128} (\bibinfo {year}
  {2009})}\BibitemShut {NoStop}%
\bibitem [{\citenamefont {Timme}\ and\ \citenamefont
  {Casadiego}(2014)}]{timme2014revealing}%
  \BibitemOpen
  \bibfield  {author} {\bibinfo {author} {\bibfnamefont {M.}~\bibnamefont
  {Timme}}\ and\ \bibinfo {author} {\bibfnamefont {J.}~\bibnamefont
  {Casadiego}},\ }\href@noop {} {\bibfield  {journal} {\bibinfo  {journal}
  {Journal of Physics A: Mathematical and Theoretical}\ }\textbf {\bibinfo
  {volume} {47}},\ \bibinfo {pages} {343001} (\bibinfo {year}
  {2014})}\BibitemShut {NoStop}%
\bibitem [{\citenamefont {Pereda}\ \emph {et~al.}(2005)\citenamefont {Pereda},
  \citenamefont {Quiroga},\ and\ \citenamefont
  {Bhattacharya}}]{pereda2005nonlinear}%
  \BibitemOpen
  \bibfield  {author} {\bibinfo {author} {\bibfnamefont {E.}~\bibnamefont
  {Pereda}}, \bibinfo {author} {\bibfnamefont {R.~Q.}\ \bibnamefont {Quiroga}},
  \ and\ \bibinfo {author} {\bibfnamefont {J.}~\bibnamefont {Bhattacharya}},\
  }\href@noop {} {\bibfield  {journal} {\bibinfo  {journal} {Progress in
  neurobiology}\ }\textbf {\bibinfo {volume} {77}},\ \bibinfo {pages} {1}
  (\bibinfo {year} {2005})}\BibitemShut {NoStop}%
\bibitem [{\citenamefont {Friston}\ \emph {et~al.}(2003)\citenamefont
  {Friston}, \citenamefont {Harrison},\ and\ \citenamefont
  {Penny}}]{friston2003dynamic}%
  \BibitemOpen
  \bibfield  {author} {\bibinfo {author} {\bibfnamefont {K.~J.}\ \bibnamefont
  {Friston}}, \bibinfo {author} {\bibfnamefont {L.}~\bibnamefont {Harrison}}, \
  and\ \bibinfo {author} {\bibfnamefont {W.}~\bibnamefont {Penny}},\
  }\href@noop {} {\bibfield  {journal} {\bibinfo  {journal} {Neuroimage}\
  }\textbf {\bibinfo {volume} {19}},\ \bibinfo {pages} {1273} (\bibinfo {year}
  {2003})}\BibitemShut {NoStop}%
\bibitem [{\citenamefont {Schelter}\ \emph {et~al.}(2006)\citenamefont
  {Schelter}, \citenamefont {Winterhalder}, \citenamefont {Eichler},
  \citenamefont {Peifer}, \citenamefont {Hellwig}, \citenamefont {Guschlbauer},
  \citenamefont {L{\"u}cking}, \citenamefont {Dahlhaus},\ and\ \citenamefont
  {Timmer}}]{schelter2006testing}%
  \BibitemOpen
  \bibfield  {author} {\bibinfo {author} {\bibfnamefont {B.}~\bibnamefont
  {Schelter}}, \bibinfo {author} {\bibfnamefont {M.}~\bibnamefont
  {Winterhalder}}, \bibinfo {author} {\bibfnamefont {M.}~\bibnamefont
  {Eichler}}, \bibinfo {author} {\bibfnamefont {M.}~\bibnamefont {Peifer}},
  \bibinfo {author} {\bibfnamefont {B.}~\bibnamefont {Hellwig}}, \bibinfo
  {author} {\bibfnamefont {B.}~\bibnamefont {Guschlbauer}}, \bibinfo {author}
  {\bibfnamefont {C.~H.}\ \bibnamefont {L{\"u}cking}}, \bibinfo {author}
  {\bibfnamefont {R.}~\bibnamefont {Dahlhaus}}, \ and\ \bibinfo {author}
  {\bibfnamefont {J.}~\bibnamefont {Timmer}},\ }\href@noop {} {\bibfield
  {journal} {\bibinfo  {journal} {Journal of neuroscience methods}\ }\textbf
  {\bibinfo {volume} {152}},\ \bibinfo {pages} {210} (\bibinfo {year}
  {2006})}\BibitemShut {NoStop}%
\bibitem [{\citenamefont {Staniek}\ and\ \citenamefont
  {Lehnertz}(2008)}]{staniek2008symbolic}%
  \BibitemOpen
  \bibfield  {author} {\bibinfo {author} {\bibfnamefont {M.}~\bibnamefont
  {Staniek}}\ and\ \bibinfo {author} {\bibfnamefont {K.}~\bibnamefont
  {Lehnertz}},\ }\href@noop {} {\bibfield  {journal} {\bibinfo  {journal}
  {Physical review letters}\ }\textbf {\bibinfo {volume} {100}},\ \bibinfo
  {pages} {158101} (\bibinfo {year} {2008})}\BibitemShut {NoStop}%
\bibitem [{\citenamefont {Andrzejak}\ and\ \citenamefont
  {Kreuz}(2011)}]{andrzejak2011characterizing}%
  \BibitemOpen
  \bibfield  {author} {\bibinfo {author} {\bibfnamefont {R.~G.}\ \bibnamefont
  {Andrzejak}}\ and\ \bibinfo {author} {\bibfnamefont {T.}~\bibnamefont
  {Kreuz}},\ }\href@noop {} {\bibfield  {journal} {\bibinfo  {journal} {EPL
  (Europhysics Letters)}\ }\textbf {\bibinfo {volume} {96}},\ \bibinfo {pages}
  {50012} (\bibinfo {year} {2011})}\BibitemShut {NoStop}%
\bibitem [{\citenamefont {Wu}\ \emph {et~al.}(2008)\citenamefont {Wu},
  \citenamefont {Liu},\ and\ \citenamefont {Feng}}]{wu2008detecting}%
  \BibitemOpen
  \bibfield  {author} {\bibinfo {author} {\bibfnamefont {J.}~\bibnamefont
  {Wu}}, \bibinfo {author} {\bibfnamefont {X.}~\bibnamefont {Liu}}, \ and\
  \bibinfo {author} {\bibfnamefont {J.}~\bibnamefont {Feng}},\ }\href@noop {}
  {\bibfield  {journal} {\bibinfo  {journal} {Journal of Neuroscience Methods}\
  }\textbf {\bibinfo {volume} {167}},\ \bibinfo {pages} {367} (\bibinfo {year}
  {2008})}\BibitemShut {NoStop}%
\bibitem [{\citenamefont {Marschinski}\ and\ \citenamefont
  {Kantz}(2002)}]{marschinski2002analysing}%
  \BibitemOpen
  \bibfield  {author} {\bibinfo {author} {\bibfnamefont {R.}~\bibnamefont
  {Marschinski}}\ and\ \bibinfo {author} {\bibfnamefont {H.}~\bibnamefont
  {Kantz}},\ }\href@noop {} {\bibfield  {journal} {\bibinfo  {journal} {The
  European Physical Journal B-Condensed Matter and Complex Systems}\ }\textbf
  {\bibinfo {volume} {30}},\ \bibinfo {pages} {275} (\bibinfo {year}
  {2002})}\BibitemShut {NoStop}%
\bibitem [{\citenamefont {Lee}(2012)}]{lee2012jumps}%
  \BibitemOpen
  \bibfield  {author} {\bibinfo {author} {\bibfnamefont {S.~S.}\ \bibnamefont
  {Lee}},\ }\href@noop {} {\bibfield  {journal} {\bibinfo  {journal} {The
  Review of Financial Studies}\ }\textbf {\bibinfo {volume} {25}},\ \bibinfo
  {pages} {439} (\bibinfo {year} {2012})}\BibitemShut {NoStop}%
\bibitem [{\citenamefont {Tissot}\ \emph {et~al.}(2014)\citenamefont {Tissot},
  \citenamefont {Lozano-Dur{\'a}n}, \citenamefont {Cordier}, \citenamefont
  {Jim{\'e}nez},\ and\ \citenamefont {Noack}}]{tissot2014granger}%
  \BibitemOpen
  \bibfield  {author} {\bibinfo {author} {\bibfnamefont {G.}~\bibnamefont
  {Tissot}}, \bibinfo {author} {\bibfnamefont {A.}~\bibnamefont
  {Lozano-Dur{\'a}n}}, \bibinfo {author} {\bibfnamefont {L.}~\bibnamefont
  {Cordier}}, \bibinfo {author} {\bibfnamefont {J.}~\bibnamefont
  {Jim{\'e}nez}}, \ and\ \bibinfo {author} {\bibfnamefont {B.~R.}\ \bibnamefont
  {Noack}},\ }in\ \href@noop {} {\emph {\bibinfo {booktitle} {Journal of
  Physics: Conference Series}}},\ Vol.\ \bibinfo {volume} {506}\ (\bibinfo
  {organization} {IOP Publishing},\ \bibinfo {year} {2014})\ p.\ \bibinfo
  {pages} {012006}\BibitemShut {NoStop}%
\bibitem [{\citenamefont {Materassi}\ \emph {et~al.}(2014)\citenamefont
  {Materassi}, \citenamefont {Consolini}, \citenamefont {Smith},\ and\
  \citenamefont {De~Marco}}]{materassi2014information}%
  \BibitemOpen
  \bibfield  {author} {\bibinfo {author} {\bibfnamefont {M.}~\bibnamefont
  {Materassi}}, \bibinfo {author} {\bibfnamefont {G.}~\bibnamefont
  {Consolini}}, \bibinfo {author} {\bibfnamefont {N.}~\bibnamefont {Smith}}, \
  and\ \bibinfo {author} {\bibfnamefont {R.}~\bibnamefont {De~Marco}},\
  }\href@noop {} {\bibfield  {journal} {\bibinfo  {journal} {Entropy}\ }\textbf
  {\bibinfo {volume} {16}},\ \bibinfo {pages} {1272} (\bibinfo {year}
  {2014})}\BibitemShut {NoStop}%
\bibitem [{\citenamefont {Vastano}\ and\ \citenamefont
  {Swinney}(1988)}]{vastano1988information}%
  \BibitemOpen
  \bibfield  {author} {\bibinfo {author} {\bibfnamefont {J.~A.}\ \bibnamefont
  {Vastano}}\ and\ \bibinfo {author} {\bibfnamefont {H.~L.}\ \bibnamefont
  {Swinney}},\ }\href@noop {} {\bibfield  {journal} {\bibinfo  {journal}
  {Physical Review Letters}\ }\textbf {\bibinfo {volume} {60}},\ \bibinfo
  {pages} {1773} (\bibinfo {year} {1988})}\BibitemShut {NoStop}%
\bibitem [{\citenamefont {Sun}\ and\ \citenamefont
  {Bollt}(2014)}]{sun2014causation}%
  \BibitemOpen
  \bibfield  {author} {\bibinfo {author} {\bibfnamefont {J.}~\bibnamefont
  {Sun}}\ and\ \bibinfo {author} {\bibfnamefont {E.~M.}\ \bibnamefont
  {Bollt}},\ }\href@noop {} {\bibfield  {journal} {\bibinfo  {journal} {Physica
  D: Nonlinear Phenomena}\ }\textbf {\bibinfo {volume} {267}},\ \bibinfo
  {pages} {49} (\bibinfo {year} {2014})}\BibitemShut {NoStop}%
\bibitem [{\citenamefont {Duan}\ \emph {et~al.}(2013)\citenamefont {Duan},
  \citenamefont {Yang}, \citenamefont {Chen},\ and\ \citenamefont
  {Shah}}]{duan2013direct}%
  \BibitemOpen
  \bibfield  {author} {\bibinfo {author} {\bibfnamefont {P.}~\bibnamefont
  {Duan}}, \bibinfo {author} {\bibfnamefont {F.}~\bibnamefont {Yang}}, \bibinfo
  {author} {\bibfnamefont {T.}~\bibnamefont {Chen}}, \ and\ \bibinfo {author}
  {\bibfnamefont {S.~L.}\ \bibnamefont {Shah}},\ }\href@noop {} {\bibfield
  {journal} {\bibinfo  {journal} {IEEE transactions on control systems
  technology}\ }\textbf {\bibinfo {volume} {21}},\ \bibinfo {pages} {2052}
  (\bibinfo {year} {2013})}\BibitemShut {NoStop}%
\bibitem [{\citenamefont {Hahs}\ and\ \citenamefont
  {Pethel}(2011)}]{hahs2011distinguishing}%
  \BibitemOpen
  \bibfield  {author} {\bibinfo {author} {\bibfnamefont {D.~W.}\ \bibnamefont
  {Hahs}}\ and\ \bibinfo {author} {\bibfnamefont {S.~D.}\ \bibnamefont
  {Pethel}},\ }\href@noop {} {\bibfield  {journal} {\bibinfo  {journal}
  {Physical review letters}\ }\textbf {\bibinfo {volume} {107}},\ \bibinfo
  {pages} {128701} (\bibinfo {year} {2011})}\BibitemShut {NoStop}%
\bibitem [{\citenamefont {Smirnov}(2013)}]{smirnov2013spurious}%
  \BibitemOpen
  \bibfield  {author} {\bibinfo {author} {\bibfnamefont {D.~A.}\ \bibnamefont
  {Smirnov}},\ }\href@noop {} {\bibfield  {journal} {\bibinfo  {journal}
  {Physical Review E}\ }\textbf {\bibinfo {volume} {87}},\ \bibinfo {pages}
  {042917} (\bibinfo {year} {2013})}\BibitemShut {NoStop}%
\bibitem [{\citenamefont {San~Liang}\ and\ \citenamefont
  {Kleeman}(2005)}]{san2005information}%
  \BibitemOpen
  \bibfield  {author} {\bibinfo {author} {\bibfnamefont {X.}~\bibnamefont
  {San~Liang}}\ and\ \bibinfo {author} {\bibfnamefont {R.}~\bibnamefont
  {Kleeman}},\ }\href@noop {} {\bibfield  {journal} {\bibinfo  {journal}
  {Physical review letters}\ }\textbf {\bibinfo {volume} {95}},\ \bibinfo
  {pages} {244101} (\bibinfo {year} {2005})}\BibitemShut {NoStop}%
\bibitem [{\citenamefont {San~Liang}(2008)}]{san2008information}%
  \BibitemOpen
  \bibfield  {author} {\bibinfo {author} {\bibfnamefont {X.}~\bibnamefont
  {San~Liang}},\ }\href@noop {} {\bibfield  {journal} {\bibinfo  {journal}
  {Physical Review E}\ }\textbf {\bibinfo {volume} {78}},\ \bibinfo {pages}
  {031113} (\bibinfo {year} {2008})}\BibitemShut {NoStop}%
\bibitem [{\citenamefont {San~Liang}(2014)}]{san2014unraveling}%
  \BibitemOpen
  \bibfield  {author} {\bibinfo {author} {\bibfnamefont {X.}~\bibnamefont
  {San~Liang}},\ }\href@noop {} {\bibfield  {journal} {\bibinfo  {journal}
  {Physical Review E}\ }\textbf {\bibinfo {volume} {90}},\ \bibinfo {pages}
  {052150} (\bibinfo {year} {2014})}\BibitemShut {NoStop}%
\bibitem [{\citenamefont {San~Liang}(2016)}]{san2016information}%
  \BibitemOpen
  \bibfield  {author} {\bibinfo {author} {\bibfnamefont {X.}~\bibnamefont
  {San~Liang}},\ }\href@noop {} {\bibfield  {journal} {\bibinfo  {journal}
  {Physical Review E}\ }\textbf {\bibinfo {volume} {94}},\ \bibinfo {pages}
  {052201} (\bibinfo {year} {2016})}\BibitemShut {NoStop}%
\bibitem [{\citenamefont {Liang}(2021{\natexlab{a}})}]{liang2021measuring}%
  \BibitemOpen
  \bibfield  {author} {\bibinfo {author} {\bibfnamefont {X.~S.}\ \bibnamefont
  {Liang}},\ }\href@noop {} {\bibfield  {journal} {\bibinfo  {journal} {arXiv
  preprint arXiv:2104.09290}\ } (\bibinfo {year}
  {2021}{\natexlab{a}})}\BibitemShut {NoStop}%
\bibitem [{\citenamefont {Liang}(2021{\natexlab{b}})}]{liang2021normalized}%
  \BibitemOpen
  \bibfield  {author} {\bibinfo {author} {\bibfnamefont {X.~S.}\ \bibnamefont
  {Liang}},\ }\href@noop {} {\bibfield  {journal} {\bibinfo  {journal}
  {Entropy}\ }\textbf {\bibinfo {volume} {23}},\ \bibinfo {pages} {679}
  (\bibinfo {year} {2021}{\natexlab{b}})}\BibitemShut {NoStop}%
\bibitem [{\citenamefont {Vannitsem}\ \emph {et~al.}(2019)\citenamefont
  {Vannitsem}, \citenamefont {Dalaiden},\ and\ \citenamefont
  {Goosse}}]{vannitsem2019testing}%
  \BibitemOpen
  \bibfield  {author} {\bibinfo {author} {\bibfnamefont {S.}~\bibnamefont
  {Vannitsem}}, \bibinfo {author} {\bibfnamefont {Q.}~\bibnamefont {Dalaiden}},
  \ and\ \bibinfo {author} {\bibfnamefont {H.}~\bibnamefont {Goosse}},\
  }\href@noop {} {\bibfield  {journal} {\bibinfo  {journal} {Geophysical
  Research Letters}\ }\textbf {\bibinfo {volume} {46}},\ \bibinfo {pages}
  {12125} (\bibinfo {year} {2019})}\BibitemShut {NoStop}%
\bibitem [{\citenamefont {Hristopulos}\ \emph {et~al.}(2019)\citenamefont
  {Hristopulos}, \citenamefont {Babul}, \citenamefont {Babul}, \citenamefont
  {Brucar},\ and\ \citenamefont {Virji-Babul}}]{hristopulos2019disrupted}%
  \BibitemOpen
  \bibfield  {author} {\bibinfo {author} {\bibfnamefont {D.~T.}\ \bibnamefont
  {Hristopulos}}, \bibinfo {author} {\bibfnamefont {A.}~\bibnamefont {Babul}},
  \bibinfo {author} {\bibfnamefont {S.}~\bibnamefont {Babul}}, \bibinfo
  {author} {\bibfnamefont {L.~R.}\ \bibnamefont {Brucar}}, \ and\ \bibinfo
  {author} {\bibfnamefont {N.}~\bibnamefont {Virji-Babul}},\ }\href@noop {}
  {\bibfield  {journal} {\bibinfo  {journal} {Frontiers in human neuroscience}\
  }\textbf {\bibinfo {volume} {13}},\ \bibinfo {pages} {419} (\bibinfo {year}
  {2019})}\BibitemShut {NoStop}%
\bibitem [{\citenamefont {Hagan}\ \emph {et~al.}(2019)\citenamefont {Hagan},
  \citenamefont {Wang}, \citenamefont {San~Liang},\ and\ \citenamefont
  {Dolman}}]{hagan2019time}%
  \BibitemOpen
  \bibfield  {author} {\bibinfo {author} {\bibfnamefont {D.~F.~T.}\
  \bibnamefont {Hagan}}, \bibinfo {author} {\bibfnamefont {G.}~\bibnamefont
  {Wang}}, \bibinfo {author} {\bibfnamefont {X.}~\bibnamefont {San~Liang}}, \
  and\ \bibinfo {author} {\bibfnamefont {H.~A.}\ \bibnamefont {Dolman}},\
  }\href@noop {} {\bibfield  {journal} {\bibinfo  {journal} {Journal of
  Climate}\ }\textbf {\bibinfo {volume} {32}},\ \bibinfo {pages} {7521}
  (\bibinfo {year} {2019})}\BibitemShut {NoStop}%
\bibitem [{\citenamefont {Stips}\ \emph {et~al.}(2016)\citenamefont {Stips},
  \citenamefont {Macias}, \citenamefont {Coughlan}, \citenamefont
  {Garcia-Gorriz},\ and\ \citenamefont {San~Liang}}]{stips2016causal}%
  \BibitemOpen
  \bibfield  {author} {\bibinfo {author} {\bibfnamefont {A.}~\bibnamefont
  {Stips}}, \bibinfo {author} {\bibfnamefont {D.}~\bibnamefont {Macias}},
  \bibinfo {author} {\bibfnamefont {C.}~\bibnamefont {Coughlan}}, \bibinfo
  {author} {\bibfnamefont {E.}~\bibnamefont {Garcia-Gorriz}}, \ and\ \bibinfo
  {author} {\bibfnamefont {X.}~\bibnamefont {San~Liang}},\ }\href@noop {}
  {\bibfield  {journal} {\bibinfo  {journal} {Scientific reports}\ }\textbf
  {\bibinfo {volume} {6}},\ \bibinfo {pages} {1} (\bibinfo {year}
  {2016})}\BibitemShut {NoStop}%
\bibitem [{\citenamefont {Bell}(1964)}]{bell1964einstein}%
  \BibitemOpen
  \bibfield  {author} {\bibinfo {author} {\bibfnamefont {J.~S.}\ \bibnamefont
  {Bell}},\ }\href@noop {} {\bibfield  {journal} {\bibinfo  {journal} {Physics
  Physique Fizika}\ }\textbf {\bibinfo {volume} {1}},\ \bibinfo {pages} {195}
  (\bibinfo {year} {1964})}\BibitemShut {NoStop}%
\bibitem [{\citenamefont {Freedman}\ and\ \citenamefont
  {Clauser}(1972)}]{freedman1972experimental}%
  \BibitemOpen
  \bibfield  {author} {\bibinfo {author} {\bibfnamefont {S.~J.}\ \bibnamefont
  {Freedman}}\ and\ \bibinfo {author} {\bibfnamefont {J.~F.}\ \bibnamefont
  {Clauser}},\ }\href@noop {} {\bibfield  {journal} {\bibinfo  {journal}
  {Physical Review Letters}\ }\textbf {\bibinfo {volume} {28}},\ \bibinfo
  {pages} {938} (\bibinfo {year} {1972})}\BibitemShut {NoStop}%
\bibitem [{\citenamefont {Aspect}\ \emph {et~al.}(1982)\citenamefont {Aspect},
  \citenamefont {Grangier},\ and\ \citenamefont
  {Roger}}]{aspect1982experimental}%
  \BibitemOpen
  \bibfield  {author} {\bibinfo {author} {\bibfnamefont {A.}~\bibnamefont
  {Aspect}}, \bibinfo {author} {\bibfnamefont {P.}~\bibnamefont {Grangier}}, \
  and\ \bibinfo {author} {\bibfnamefont {G.}~\bibnamefont {Roger}},\
  }\href@noop {} {\bibfield  {journal} {\bibinfo  {journal} {Physical review
  letters}\ }\textbf {\bibinfo {volume} {49}},\ \bibinfo {pages} {91} (\bibinfo
  {year} {1982})}\BibitemShut {NoStop}%
\bibitem [{\citenamefont {Christensen}\ \emph {et~al.}(2013)\citenamefont
  {Christensen}, \citenamefont {McCusker}, \citenamefont {Altepeter},
  \citenamefont {Calkins}, \citenamefont {Gerrits}, \citenamefont {Lita},
  \citenamefont {Miller}, \citenamefont {Shalm}, \citenamefont {Zhang},
  \citenamefont {Nam} \emph {et~al.}}]{christensen2013detection}%
  \BibitemOpen
  \bibfield  {author} {\bibinfo {author} {\bibfnamefont {B.~G.}\ \bibnamefont
  {Christensen}}, \bibinfo {author} {\bibfnamefont {K.~T.}\ \bibnamefont
  {McCusker}}, \bibinfo {author} {\bibfnamefont {J.~B.}\ \bibnamefont
  {Altepeter}}, \bibinfo {author} {\bibfnamefont {B.}~\bibnamefont {Calkins}},
  \bibinfo {author} {\bibfnamefont {T.}~\bibnamefont {Gerrits}}, \bibinfo
  {author} {\bibfnamefont {A.~E.}\ \bibnamefont {Lita}}, \bibinfo {author}
  {\bibfnamefont {A.}~\bibnamefont {Miller}}, \bibinfo {author} {\bibfnamefont
  {L.~K.}\ \bibnamefont {Shalm}}, \bibinfo {author} {\bibfnamefont
  {Y.}~\bibnamefont {Zhang}}, \bibinfo {author} {\bibfnamefont {S.~W.}\
  \bibnamefont {Nam}},  \emph {et~al.},\ }\href@noop {} {\bibfield  {journal}
  {\bibinfo  {journal} {Physical review letters}\ }\textbf {\bibinfo {volume}
  {111}},\ \bibinfo {pages} {130406} (\bibinfo {year} {2013})}\BibitemShut
  {NoStop}%
\bibitem [{\citenamefont {Rowe}\ \emph {et~al.}(2001)\citenamefont {Rowe},
  \citenamefont {Kielpinski}, \citenamefont {Meyer}, \citenamefont {Sackett},
  \citenamefont {Itano}, \citenamefont {Monroe},\ and\ \citenamefont
  {Wineland}}]{rowe2001experimental}%
  \BibitemOpen
  \bibfield  {author} {\bibinfo {author} {\bibfnamefont {M.~A.}\ \bibnamefont
  {Rowe}}, \bibinfo {author} {\bibfnamefont {D.}~\bibnamefont {Kielpinski}},
  \bibinfo {author} {\bibfnamefont {V.}~\bibnamefont {Meyer}}, \bibinfo
  {author} {\bibfnamefont {C.~A.}\ \bibnamefont {Sackett}}, \bibinfo {author}
  {\bibfnamefont {W.~M.}\ \bibnamefont {Itano}}, \bibinfo {author}
  {\bibfnamefont {C.}~\bibnamefont {Monroe}}, \ and\ \bibinfo {author}
  {\bibfnamefont {D.~J.}\ \bibnamefont {Wineland}},\ }\href@noop {} {\bibfield
  {journal} {\bibinfo  {journal} {Nature}\ }\textbf {\bibinfo {volume} {409}},\
  \bibinfo {pages} {791} (\bibinfo {year} {2001})}\BibitemShut {NoStop}%
\bibitem [{\citenamefont {Giustina}\ \emph {et~al.}(2013)\citenamefont
  {Giustina}, \citenamefont {Mech}, \citenamefont {Ramelow}, \citenamefont
  {Wittmann}, \citenamefont {Kofler}, \citenamefont {Beyer}, \citenamefont
  {Lita}, \citenamefont {Calkins}, \citenamefont {Gerrits}, \citenamefont {Nam}
  \emph {et~al.}}]{giustina2013bell}%
  \BibitemOpen
  \bibfield  {author} {\bibinfo {author} {\bibfnamefont {M.}~\bibnamefont
  {Giustina}}, \bibinfo {author} {\bibfnamefont {A.}~\bibnamefont {Mech}},
  \bibinfo {author} {\bibfnamefont {S.}~\bibnamefont {Ramelow}}, \bibinfo
  {author} {\bibfnamefont {B.}~\bibnamefont {Wittmann}}, \bibinfo {author}
  {\bibfnamefont {J.}~\bibnamefont {Kofler}}, \bibinfo {author} {\bibfnamefont
  {J.}~\bibnamefont {Beyer}}, \bibinfo {author} {\bibfnamefont
  {A.}~\bibnamefont {Lita}}, \bibinfo {author} {\bibfnamefont {B.}~\bibnamefont
  {Calkins}}, \bibinfo {author} {\bibfnamefont {T.}~\bibnamefont {Gerrits}},
  \bibinfo {author} {\bibfnamefont {S.~W.}\ \bibnamefont {Nam}},  \emph
  {et~al.},\ }\href@noop {} {\bibfield  {journal} {\bibinfo  {journal}
  {Nature}\ }\textbf {\bibinfo {volume} {497}},\ \bibinfo {pages} {227}
  (\bibinfo {year} {2013})}\BibitemShut {NoStop}%
\bibitem [{\citenamefont {Gachechiladze}\ \emph {et~al.}(2020)\citenamefont
  {Gachechiladze}, \citenamefont {Miklin},\ and\ \citenamefont
  {Chaves}}]{gachechiladze2020quantifying}%
  \BibitemOpen
  \bibfield  {author} {\bibinfo {author} {\bibfnamefont {M.}~\bibnamefont
  {Gachechiladze}}, \bibinfo {author} {\bibfnamefont {N.}~\bibnamefont
  {Miklin}}, \ and\ \bibinfo {author} {\bibfnamefont {R.}~\bibnamefont
  {Chaves}},\ }\href@noop {} {\bibfield  {journal} {\bibinfo  {journal}
  {Physical Review Letters}\ }\textbf {\bibinfo {volume} {125}},\ \bibinfo
  {pages} {230401} (\bibinfo {year} {2020})}\BibitemShut {NoStop}%
\bibitem [{\citenamefont {Henson}\ \emph {et~al.}(2014)\citenamefont {Henson},
  \citenamefont {Lal},\ and\ \citenamefont {Pusey}}]{henson2014theory}%
  \BibitemOpen
  \bibfield  {author} {\bibinfo {author} {\bibfnamefont {J.}~\bibnamefont
  {Henson}}, \bibinfo {author} {\bibfnamefont {R.}~\bibnamefont {Lal}}, \ and\
  \bibinfo {author} {\bibfnamefont {M.~F.}\ \bibnamefont {Pusey}},\ }\href@noop
  {} {\bibfield  {journal} {\bibinfo  {journal} {New Journal of Physics}\
  }\textbf {\bibinfo {volume} {16}},\ \bibinfo {pages} {113043} (\bibinfo
  {year} {2014})}\BibitemShut {NoStop}%
\bibitem [{\citenamefont {Chaves}\ \emph {et~al.}(2015)\citenamefont {Chaves},
  \citenamefont {Majenz},\ and\ \citenamefont {Gross}}]{chaves2015information}%
  \BibitemOpen
  \bibfield  {author} {\bibinfo {author} {\bibfnamefont {R.}~\bibnamefont
  {Chaves}}, \bibinfo {author} {\bibfnamefont {C.}~\bibnamefont {Majenz}}, \
  and\ \bibinfo {author} {\bibfnamefont {D.}~\bibnamefont {Gross}},\
  }\href@noop {} {\bibfield  {journal} {\bibinfo  {journal} {Nature
  communications}\ }\textbf {\bibinfo {volume} {6}},\ \bibinfo {pages} {1}
  (\bibinfo {year} {2015})}\BibitemShut {NoStop}%
\bibitem [{\citenamefont {Costa}\ and\ \citenamefont
  {Shrapnel}(2016)}]{costa2016quantum}%
  \BibitemOpen
  \bibfield  {author} {\bibinfo {author} {\bibfnamefont {F.}~\bibnamefont
  {Costa}}\ and\ \bibinfo {author} {\bibfnamefont {S.}~\bibnamefont
  {Shrapnel}},\ }\href@noop {} {\bibfield  {journal} {\bibinfo  {journal} {New
  Journal of Physics}\ }\textbf {\bibinfo {volume} {18}},\ \bibinfo {pages}
  {063032} (\bibinfo {year} {2016})}\BibitemShut {NoStop}%
\bibitem [{\citenamefont {Fritz}(2016)}]{fritz2016beyond}%
  \BibitemOpen
  \bibfield  {author} {\bibinfo {author} {\bibfnamefont {T.}~\bibnamefont
  {Fritz}},\ }\href@noop {} {\bibfield  {journal} {\bibinfo  {journal}
  {Communications in Mathematical Physics}\ }\textbf {\bibinfo {volume}
  {341}},\ \bibinfo {pages} {391} (\bibinfo {year} {2016})}\BibitemShut
  {NoStop}%
\bibitem [{\citenamefont {Barrett}\ \emph {et~al.}(2019)\citenamefont
  {Barrett}, \citenamefont {Lorenz},\ and\ \citenamefont
  {Oreshkov}}]{barrett2019quantum}%
  \BibitemOpen
  \bibfield  {author} {\bibinfo {author} {\bibfnamefont {J.}~\bibnamefont
  {Barrett}}, \bibinfo {author} {\bibfnamefont {R.}~\bibnamefont {Lorenz}}, \
  and\ \bibinfo {author} {\bibfnamefont {O.}~\bibnamefont {Oreshkov}},\
  }\href@noop {} {\bibfield  {journal} {\bibinfo  {journal} {arXiv preprint
  arXiv:1906.10726}\ } (\bibinfo {year} {2019})}\BibitemShut {NoStop}%
\bibitem [{\citenamefont {Wolfe}\ \emph {et~al.}(2021)\citenamefont {Wolfe},
  \citenamefont {Pozas-Kerstjens}, \citenamefont {Grinberg}, \citenamefont
  {Rosset}, \citenamefont {Ac{\'\i}n},\ and\ \citenamefont
  {Navascu{\'e}s}}]{wolfe2021quantum}%
  \BibitemOpen
  \bibfield  {author} {\bibinfo {author} {\bibfnamefont {E.}~\bibnamefont
  {Wolfe}}, \bibinfo {author} {\bibfnamefont {A.}~\bibnamefont
  {Pozas-Kerstjens}}, \bibinfo {author} {\bibfnamefont {M.}~\bibnamefont
  {Grinberg}}, \bibinfo {author} {\bibfnamefont {D.}~\bibnamefont {Rosset}},
  \bibinfo {author} {\bibfnamefont {A.}~\bibnamefont {Ac{\'\i}n}}, \ and\
  \bibinfo {author} {\bibfnamefont {M.}~\bibnamefont {Navascu{\'e}s}},\
  }\href@noop {} {\bibfield  {journal} {\bibinfo  {journal} {Physical Review
  X}\ }\textbf {\bibinfo {volume} {11}},\ \bibinfo {pages} {021043} (\bibinfo
  {year} {2021})}\BibitemShut {NoStop}%
\bibitem [{\citenamefont {{\AA}berg}\ \emph {et~al.}(2020)\citenamefont
  {{\AA}berg}, \citenamefont {Nery}, \citenamefont {Duarte},\ and\
  \citenamefont {Chaves}}]{aaberg2020semidefinite}%
  \BibitemOpen
  \bibfield  {author} {\bibinfo {author} {\bibfnamefont {J.}~\bibnamefont
  {{\AA}berg}}, \bibinfo {author} {\bibfnamefont {R.}~\bibnamefont {Nery}},
  \bibinfo {author} {\bibfnamefont {C.}~\bibnamefont {Duarte}}, \ and\ \bibinfo
  {author} {\bibfnamefont {R.}~\bibnamefont {Chaves}},\ }\href@noop {}
  {\bibfield  {journal} {\bibinfo  {journal} {Physical Review Letters}\
  }\textbf {\bibinfo {volume} {125}},\ \bibinfo {pages} {110505} (\bibinfo
  {year} {2020})}\BibitemShut {NoStop}%
\bibitem [{\citenamefont {Fitzsimons}\ \emph {et~al.}(2015)\citenamefont
  {Fitzsimons}, \citenamefont {Jones},\ and\ \citenamefont
  {Vedral}}]{fitzsimons2015quantum}%
  \BibitemOpen
  \bibfield  {author} {\bibinfo {author} {\bibfnamefont {J.~F.}\ \bibnamefont
  {Fitzsimons}}, \bibinfo {author} {\bibfnamefont {J.~A.}\ \bibnamefont
  {Jones}}, \ and\ \bibinfo {author} {\bibfnamefont {V.}~\bibnamefont
  {Vedral}},\ }\href@noop {} {\bibfield  {journal} {\bibinfo  {journal}
  {Scientific reports}\ }\textbf {\bibinfo {volume} {5}},\ \bibinfo {pages} {1}
  (\bibinfo {year} {2015})}\BibitemShut {NoStop}%
\bibitem [{\citenamefont {Pisarczyk}\ \emph {et~al.}(2019)\citenamefont
  {Pisarczyk}, \citenamefont {Zhao}, \citenamefont {Ouyang}, \citenamefont
  {Vedral},\ and\ \citenamefont {Fitzsimons}}]{pisarczyk2019causal}%
  \BibitemOpen
  \bibfield  {author} {\bibinfo {author} {\bibfnamefont {R.}~\bibnamefont
  {Pisarczyk}}, \bibinfo {author} {\bibfnamefont {Z.}~\bibnamefont {Zhao}},
  \bibinfo {author} {\bibfnamefont {Y.}~\bibnamefont {Ouyang}}, \bibinfo
  {author} {\bibfnamefont {V.}~\bibnamefont {Vedral}}, \ and\ \bibinfo {author}
  {\bibfnamefont {J.~F.}\ \bibnamefont {Fitzsimons}},\ }\href@noop {}
  {\bibfield  {journal} {\bibinfo  {journal} {Physical review letters}\
  }\textbf {\bibinfo {volume} {123}},\ \bibinfo {pages} {150502} (\bibinfo
  {year} {2019})}\BibitemShut {NoStop}%
\bibitem [{\citenamefont {Di~Franco}\ \emph {et~al.}(2007)\citenamefont
  {Di~Franco}, \citenamefont {Paternostro}, \citenamefont {Palma},\ and\
  \citenamefont {Kim}}]{di2007information}%
  \BibitemOpen
  \bibfield  {author} {\bibinfo {author} {\bibfnamefont {C.}~\bibnamefont
  {Di~Franco}}, \bibinfo {author} {\bibfnamefont {M.}~\bibnamefont
  {Paternostro}}, \bibinfo {author} {\bibfnamefont {G.}~\bibnamefont {Palma}},
  \ and\ \bibinfo {author} {\bibfnamefont {M.}~\bibnamefont {Kim}},\
  }\href@noop {} {\bibfield  {journal} {\bibinfo  {journal} {Physical Review
  A}\ }\textbf {\bibinfo {volume} {76}},\ \bibinfo {pages} {042316} (\bibinfo
  {year} {2007})}\BibitemShut {NoStop}%
\bibitem [{\citenamefont {Di~Franco}\ \emph {et~al.}(2008)\citenamefont
  {Di~Franco}, \citenamefont {Paternostro},\ and\ \citenamefont
  {Palma}}]{di2008deeper}%
  \BibitemOpen
  \bibfield  {author} {\bibinfo {author} {\bibfnamefont {C.}~\bibnamefont
  {Di~Franco}}, \bibinfo {author} {\bibfnamefont {M.}~\bibnamefont
  {Paternostro}}, \ and\ \bibinfo {author} {\bibfnamefont {G.}~\bibnamefont
  {Palma}},\ }\href@noop {} {\bibfield  {journal} {\bibinfo  {journal}
  {International Journal of Quantum Information}\ }\textbf {\bibinfo {volume}
  {6}},\ \bibinfo {pages} {659} (\bibinfo {year} {2008})}\BibitemShut {NoStop}%
\bibitem [{\citenamefont {Nielsen}\ and\ \citenamefont
  {Chuang}(2002)}]{nielsen2002quantum}%
  \BibitemOpen
  \bibfield  {author} {\bibinfo {author} {\bibfnamefont {M.~A.}\ \bibnamefont
  {Nielsen}}\ and\ \bibinfo {author} {\bibfnamefont {I.}~\bibnamefont
  {Chuang}},\ }\href@noop {} {\enquote {\bibinfo {title} {Quantum computation
  and quantum information},}\ } (\bibinfo {year} {2002})\BibitemShut {NoStop}%
\bibitem [{\citenamefont {Yung}\ \emph {et~al.}(2003)\citenamefont {Yung},
  \citenamefont {Leung},\ and\ \citenamefont {Bose}}]{yung2003exact}%
  \BibitemOpen
  \bibfield  {author} {\bibinfo {author} {\bibfnamefont {M.-H.}\ \bibnamefont
  {Yung}}, \bibinfo {author} {\bibfnamefont {D.~W.}\ \bibnamefont {Leung}}, \
  and\ \bibinfo {author} {\bibfnamefont {S.}~\bibnamefont {Bose}},\ }\href@noop
  {} {\bibfield  {journal} {\bibinfo  {journal} {arXiv preprint
  quant-ph/0312105}\ } (\bibinfo {year} {2003})}\BibitemShut {NoStop}%
\bibitem [{\citenamefont {Benjamin}\ and\ \citenamefont
  {Bose}(2004)}]{benjamin2004quantum}%
  \BibitemOpen
  \bibfield  {author} {\bibinfo {author} {\bibfnamefont {S.~C.}\ \bibnamefont
  {Benjamin}}\ and\ \bibinfo {author} {\bibfnamefont {S.}~\bibnamefont
  {Bose}},\ }\href@noop {} {\bibfield  {journal} {\bibinfo  {journal} {Physical
  Review A}\ }\textbf {\bibinfo {volume} {70}},\ \bibinfo {pages} {032314}
  (\bibinfo {year} {2004})}\BibitemShut {NoStop}%
\bibitem [{\citenamefont {Franco}\ \emph {et~al.}(2013)\citenamefont {Franco},
  \citenamefont {Bellomo}, \citenamefont {Maniscalco},\ and\ \citenamefont
  {Compagno}}]{franco2013dynamics}%
  \BibitemOpen
  \bibfield  {author} {\bibinfo {author} {\bibfnamefont {R.~L.}\ \bibnamefont
  {Franco}}, \bibinfo {author} {\bibfnamefont {B.}~\bibnamefont {Bellomo}},
  \bibinfo {author} {\bibfnamefont {S.}~\bibnamefont {Maniscalco}}, \ and\
  \bibinfo {author} {\bibfnamefont {G.}~\bibnamefont {Compagno}},\ }\href@noop
  {} {\bibfield  {journal} {\bibinfo  {journal} {International Journal of
  Modern Physics B}\ }\textbf {\bibinfo {volume} {27}},\ \bibinfo {pages}
  {1345053} (\bibinfo {year} {2013})}\BibitemShut {NoStop}%
\end{thebibliography}%

%

\appendix
\section{Brief review of classical information flow-based causality analysis}\label{classicalLianginfo}
Liang information flow quantitatively defines causality. The series of work starts with the investigation of bi-variate deterministic systems and is originally based on a heuristic argument\cite{san2005information}. Later on, the formalism is put on a rigorous footing and generalized to stochastic and multi-variate systems\cite{san2008information,san2016information,liang2021normalized}. To present this fundamental idea in its simplest form, we will focus on  bi-variate autonomous system with equation of motion given by: 
\begin{equation}
\frac{d \mathbf{x}}{dt}=\mathbf{F}(\mathbf{x})\label{eom} 
\end{equation} 
where $\mathbf{x}=(x_1,x_2)\in \Omega$ and the sample space $\Omega$ is a direct product of subspace $\Omega_1 \otimes \Omega_2$. $\mathbf{X}=(X_1,X_2)$ is the random variable of subsystem 1 and 2. $\{\mathbf{X},t\}$ is assumed a stochastic process and the joint probability density distribution at time t is denoted $\rho(x_1,x_2,t)$. $\mathbf{F}=(F_1,F_2)$ may be interpreted as the force acting on the system. Shannon entropy of this system is given by: 
\begin{equation}
S_{(classical)}(t)=-\int_{\Omega} \rho log(\rho) dx_1 dx_2 \label{classical entropy}
\end{equation}
Substitute eq\ref{eom} into eq\ref{classical entropy}, one obtains the time rate change of entropy, provided that $\rho$ vanishes at boundaries\cite{san2005information}:
\begin{equation}
\frac{dS_{(classical)}}{dt}=E(\mathbf{\nabla} \cdot \mathbf{F})\label{dS}
\end{equation}  
The right hand side is the expectation value of the divergence of force $\mathbf{F}$. The physics revealed by eq\ref{dS} is that the expansion and contraction of the phase space governs the change of entropy.  

The probability distribution of a subsystem, say subsystem 1, can be obtained by taking the marginal density $\rho_1(x_1,t)=\int_{\Omega_2}\rho(x_1,x_2,t)dx_2$. Its entropy can be calculated:
\begin{equation}
\frac{dS_{1(classical)}}{dt}=-\int_{\Omega}\rho[\frac{F_1}{\rho_1}\frac{\partial \rho_1}{\partial x_1}]dx_1dx_2 \label{dS1}
\end{equation}
Liang and Kleeman identified that the entropy change of subsystem 1 given by eq\ref{dS1} can be decomposed into two parts: the evolution due to $X_1$ alone, with effect from subsystem 2 excluded, denoted as $\frac{dS_{1\not{2}(classical)}}{dt}$. Another part is the influence from $X_2$ through the coupling with external force. Through heuristic reasoning based on the interpretation of eq\ref{dS}, Liang and Kleeman argue that if subsystem 1 evolves on its own, the entropy change of subsystem 1 would depend only on $\partial F_1/\partial x_1$:
\begin{equation}
\frac{dS_{1\not{2}(classical)}}{dt}=E(\frac{\partial F_1}{\partial x_1})=\int_{\Omega}\rho \frac{\partial F_1}{\partial x_1} dx_1 dx_2  \label{1not2}
\end{equation}
Later on, Liang(2016\cite{san2016information}) proved that the above result eq\ref{1not2} can be derived by treating $x_2$ as a fixed parameter at time $t$, rather than a variate. 

The rate of information flow from $X_2$ to $X_1$ is then:
\begin{eqnarray}
T_{2\rightarrow 1}&=&\frac{dS_{1(classical)}}{dt}-\frac{dS_{1\not{2}(classical)}}{dt}\nonumber \\
&=&-\int_{\Omega}\rho[\frac{F_1}{\rho_1}\frac{\partial \rho_1}{\partial x_1}+\frac{\partial F_1}{\partial x_1}]dx_1dx_2 \label{Tc}
\end{eqnarray}
This formula verifies what Liang refers to as \textit{the principle of nil causality}:

\vspace{1cm}
\emph{If $F_1$ is independent of $x_2$, then the information flow from 2 to 1 vanishes: $T_{2\rightarrow 1}=0$.}  
\vspace{1cm}

If $T_{2\rightarrow 1}$ is negative (positive), the interpretation is that system 2 is making system 1 more (less) certain. Note that the information flow formalism eq\ref{Tc} is asymmetric, that is $T_{2\rightarrow 1}\neq T_{1\rightarrow 2}$. When the information flow from 2 to 1 vanishes, that from 1 to 2 maybe non-zero. The asymmetry feature distinguishes the information flow formalism with classical correlation measures.  

It should be pointed out that the evaluation of eq\ref{Tc} requires full knowledge of the dynamics. In 2014\cite{san2014unraveling}, Liang showed that $T_{2\rightarrow 1}$ can be estimated with local statistics. The maximum-likelihood estimator of eq\ref{Tc} is shown to be a combination of some sample covariances, which greatly facilitates the implementation of the causality analysis.

This formalism has been widely applied to realistic schemes\cite{san2014unraveling,stips2016causal,hagan2019time,vannitsem2019testing,hristopulos2019disrupted}. Among them, we will briefly mention its application to a network consisting of Stuart-Landau oscillators\cite{liang2021measuring}, a typical model for many biological phenomena\cite{strogatz2001exploring}. The magnitude of Liang information flow quantifies the influence of individual components to produce the collective behavior of the whole system. The direct addition of individual contributions does not equal the cumulative information flow, demonstrating its collective property. Moreover, the node with greatest information flow is verified to be the most crucial as its suppression leads to shut down of the entire network. Surprisingly, such a node may be sparsely connected, rather than a center of network. The information-flow based causality analysis successfully explains why small defects at local node could severely damage structural integrity. 

\section{Classical closed bivariate system}\label{classicalbipartite}
The classical model considered in eq\ref{eom} is dissipative. System 1 and 2 exchanges energy with the environment through external force $\mathbf{F}$. If system 1 and 2 is closed, the divergence of force $\mathbf{F}$ vanishes: $\mathbf{\nabla} \cdot \mathbf{F}=0$. As a result, eq\ref{dS}, eq\ref{1not2} becomes: $dS_{(classical)}/dt=E(\mathbf{\nabla} \cdot \mathbf{F})=0$, $dS_{1\not{2}(classical)}/dt=E(\frac{\partial F_1}{\partial x_1})=0$, therefore,
\begin{equation}
T_{2\rightarrow 1}=\frac{dS_{1(classical)}}{dt}\label{eq10}
\end{equation}
Eq\ref{eq10} is completely in agreement with the quantum formalism obtained for initially mixed bipartite system.

\section{\textit{the principle of nil causality in Quantum regime}}\label{causality}
For tripartite system, if $U_{A\not{B}C}(t)$ takes the form of eq(4) in the main text ($\mathcal{V}_{AC}\otimes \mathcal{W}_B$), then the statement of causality is satisfied, that is, $T_{B\rightarrow A}=0$ when A evolves independent of B. 
\begin{proof}
	If $U_{ABC}=\mathcal{M}_A \otimes \mathcal{N}_{BC}$, the evolution of A is solely determined by unitary operator $\mathcal{M}_A$. Excluding B from the joint evolution of subsystem BC, denoted $\mathcal{N}_{\not{B}C}$, has no effect on A. Therefore, $\rho_A(t)=\rho_{A\not{B}}(t)=\mathcal{M}_A \rho_A(0) \mathcal{M}_A^\dagger$. By the unitary invariance of von-Neumann entropy, $\frac{dS_A}{dt}=\frac{dS_{A\not{B}}}{dt}=0$, thus $T_{B\rightarrow A}=0$.
	
	If $U_{ABC}(t)=\mathcal{O}_{AC} \otimes \mathcal{Q}_{B}$, it is already of the form given in eq\ref{UnotB}. Therefore, excluding B or not has no impact on the joint evolution of system AC. That is, 
	\begin{eqnarray}
	\rho_A(t)&=&\mathrm{Tr}_{BC}\{U_{ABC}(t)\rho_{ABC}(0 )U^\dagger_{ABC}(t)\}\nonumber\\
	&=&\mathrm{Tr}_C[\mathcal{O}_{AC} \rho_{AC}(0) \mathcal{O}_{AC}^\dagger]\nonumber\\
	&=&\mathrm{Tr}_{BC}\{U_{A\not{B}C}(t)\rho_{ABC}(0 )U^\dagger_{A\not{B}C}(t)\}=\rho_{A\not{B}}(t)\nonumber
	\end{eqnarray}
	Therefore, $T_{B\rightarrow A}=\frac{dS_A}{dt}-\frac{dS_{A\not{B}}}{dt}=0$. 
\end{proof}
This results obtained above can be easily extended to multi-dimensions. Whether the converse proof also holds remains an open question.

\section{Relative coupling strength variation for 5 qubits}\label{5qubit}
For 5 qubits, labeled A,B,C,D,E, with E in the center and interacting with other qubits independently. To check if stronger coupled sending qubit delivers more information towards the receiving qubit, we set $\eta_{DE}=1$, $\eta_{CE}=2$, $\eta_{BE}=3$, $\eta_{AE}=4$ and let the initial state of the sending qubits A,B,C,D being maximally mixed and the receiving qubit E pure, so that $\rho_0=I_A/2\otimes I_B/2\otimes I_C/2\otimes I_D/2\otimes |0\rangle \langle 0|_E$. 

\begin{figure}[h!]
	\includegraphics[width=1
	\columnwidth]{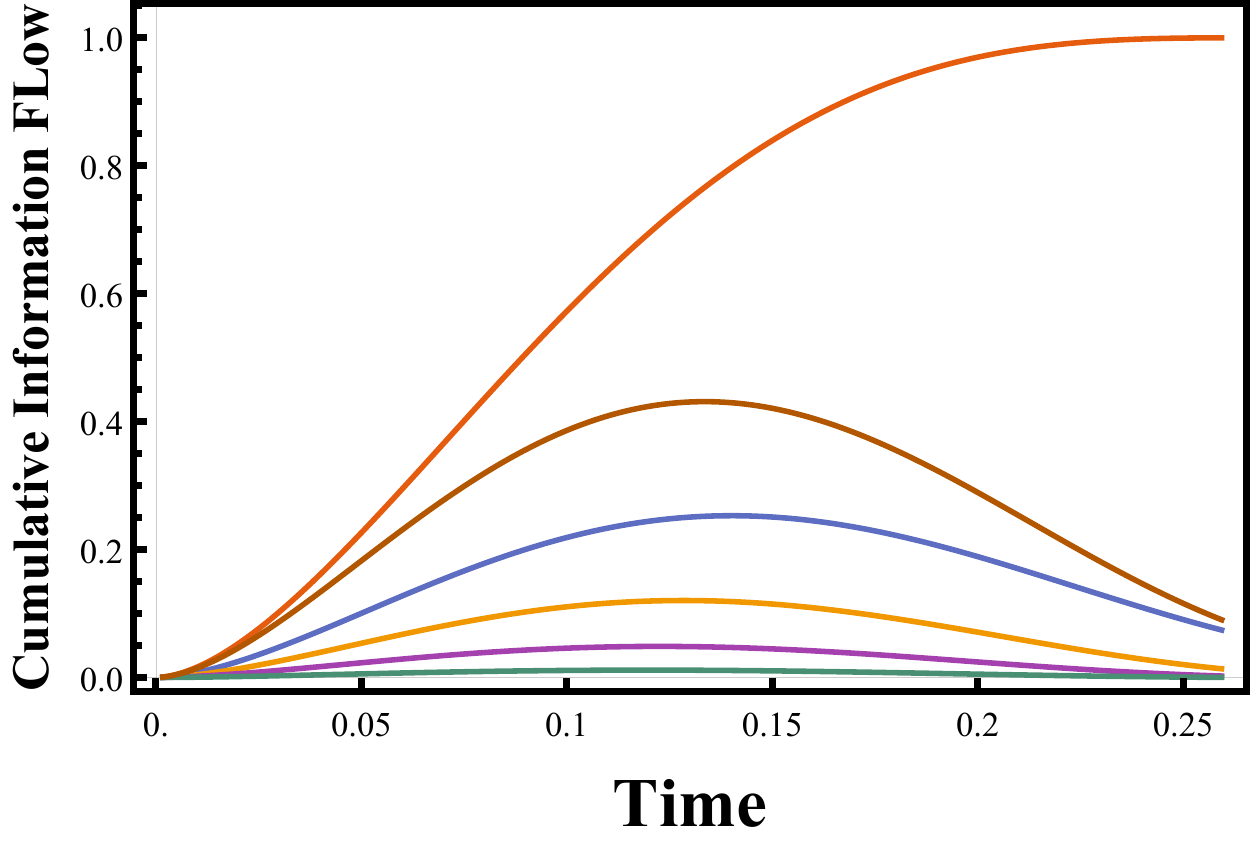}
	\caption{\footnotesize \textbf{Cumulative Information flow towards qubit E (in Bits)} from top to bottom: $\mathbb{T}_{ABCD\rightarrow E}$, $\mathbb{T}_{A\rightarrow E}+\mathbb{T}_{B\rightarrow E}+\mathbb{T}_{C\rightarrow E}+\mathbb{T}_{D \rightarrow E}$,$\mathbb{T}_{A\rightarrow E}$, $\mathbb{T}_{B\rightarrow E}$, $\mathbb{T}_{C\rightarrow E}$, $\mathbb{T}_{D\rightarrow E}$}\label{fig4} 
\end{figure}

Calculation of information flow from the $k^{th}$ qubit to E, where k runs through the sending qubits, requires the evolution mechanism with the $k^{th}$ qubit frozen:
\begin{equation}
H_{spin,\not{k}}=\sum_{i,i\neq k}H_{spin,iE}
\end{equation}
The joint information flow from A,B,C,D to E is simply the change of $S_E$:
\begin{equation}
\mathbb{T}_{ABCD\rightarrow E}=\Delta S_E \label{ABCDtoE}
\end{equation}
At time $t\sim0.26$, the entropy of E reaches its maxima $S_E=1bit$ for the first time. The Information flow from each sending qubit to E is plotted in figure \ref{fig4}, before the capacity is reached. The stronger coupled qubit delivers more information to E at all time during $t\in [0,0.26]$:
\begin{equation}
\mathbb{T}_{A\rightarrow E}>\mathbb{T}_{B\rightarrow E}>\mathbb{T}_{C\rightarrow E}>\mathbb{T}_{D \rightarrow E}
\end{equation}
At $t=0.26$, $\mathbb{T}_{A\rightarrow E}\sim 0.0731$bits, $\mathbb{T}_{B\rightarrow E}\sim 0.0132$bits, $\mathbb{T}_{C\rightarrow E}\sim 0.0022$bits,  $\mathbb{T}_{D\rightarrow E}\sim 0.0001$bits. Similar to the results obtained for 3 qubit system in the main text, here we also observe superadditivity of quantum Liang information flow:
\begin{equation}
\mathbb{T}_{ABCD\rightarrow E}>\mathbb{T}_{A\rightarrow E}+\mathbb{T}_{B\rightarrow E}+\mathbb{T}_{C\rightarrow E}+\mathbb{T}_{D \rightarrow E}
\end{equation}
\section{Schematic diagram of a 5-qubit network}
With the Hamiltonian given by eq(12) in the main text, the corresponding schematic diagram is shown below:
\begin{figure}[h]
	\includegraphics[width=1 
	\columnwidth]{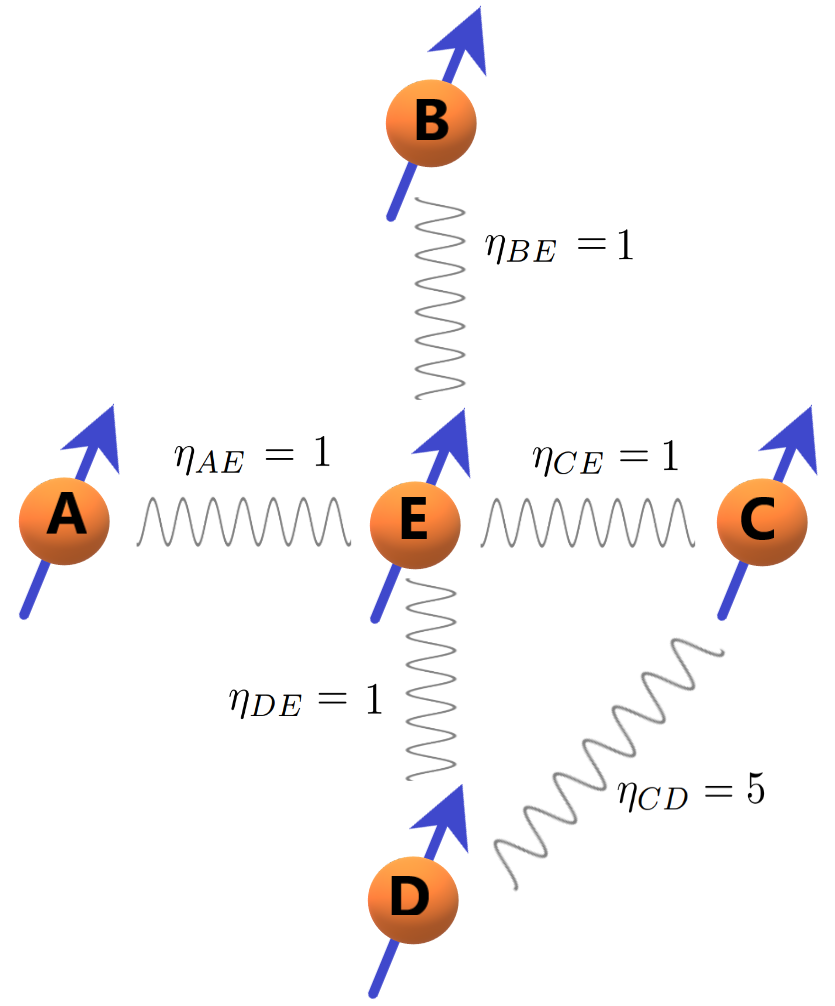}
	\caption{\footnotesize \textbf{schematic diagram:} A,B couples solely with E, while C,D also interacts with each other. $\eta_{DE}=\eta_{CE}=\eta_{BE}=\eta_{AE}=1$, $\eta_{CD}=5$} \label{fig7} 
\end{figure}
\newpage

\end{document}